\documentclass[showpacs,aps,pra,twocolumn,groupedaddress,longbibliography]{revtex4-1}

\usepackage{threeparttable}
\usepackage{setspace}
\usepackage{graphicx}
\usepackage{amsmath,mathtools,amsxtra,amssymb}
\usepackage{extarrows}
\usepackage{placeins}
\usepackage[thinlines,thicklines]{easybmat}
\accentedsymbol\Cm{\mathbf{C}}
\accentedsymbol\Lm{\mathbf{L}}
\accentedsymbol\Cmi{\mathbf{C}^{-1}}
\accentedsymbol\Lmi{\mathbf{L}^{-1}}
\accentedsymbol\Cbar{\mathbf{\overline{C}}}
\accentedsymbol\Cbbar{\mathbf{\overline{\overline{C}}}}
\accentedsymbol\Lbar{\mathbf{\overline{L}}}
\accentedsymbol\Lbbar{\mathbf{\overline{\overline{L}}}}

\begin{document}


\title{Efficient numerical simulation of complex Josephson quantum circuits}

\author{Andrew J. Kerman}
\affiliation{Lincoln Laboratory, Massachusetts Institute of
Technology, Lexington, MA, 02420}

\date{\today}

\begin{abstract}
Building on the established methods for superconducting circuit quantization, we present a new theoretical framework for approximate numerical simulation of Josephson quantum circuits. Simulations based on this framework provide access to a degree of complexity and circuit size heretofore inaccessible to quantitative analysis, including fundamentally new kinds of superconducting quantum devices. This capability is made possible by two improvements over previous methods: first, physically-motivated choices for the canonical circuit modes and physical basis states which allow a highly-efficient matrix representation; and second, an iterative method in which subsystems are diagonalized separately and then coupled together, at increasing size scales with each iteration, allowing diagonalization of Hamiltonians in extremely large Hilbert spaces to be approximated using a sequence of diagonalizations in much smaller spaces. 
\end{abstract}

\maketitle

\section{Introduction}

Superconducting quantum circuits built on the nonlinearity of Josephson junctions present an extraordinarily large design space, of which present-day device research has very likely only scratched the surface. Because of the complexity of this design space, the vast majority of work in this area has been largely built on the foundation of a few, relatively simple devices amenable to numerical simulation or semi-analytic treatment. A case in point is the ubiquitous circuit-QED architecture ~\cite{blaisCQED}, which consists of transmon qubits ~\cite{transmon} (weakly-anharmonic oscillators) coupled dispersively to microwave resonators. In these systems, both the Josephson nonlinearities and the couplings between subcircuits are weak and can be treated perturbatively, allowing analytic approximations for many important properties, even for systems involving several of these qubits. In another, parallel line of research, so-called quantum annealing systems ~\cite{QARMP} have been developed, based on flux qubits ~\cite{orlando,Dwaveflux,MITLLflux,MITLLcoupled,googleflux}, which naturally emulate transverse field Ising spin models ~\cite{Dwave3DTFIM,DwaveKT}. Although the flux qubit intrinsically relies on a stronger Josephson nonlinearity than the transmon, treatment of these systems has also been dramatically simplified by the fact that their low impedance allows a natural, semiclassical treatment of the strong nonlinearity, again permitting semi-analytic results to be obtained \cite{vandenBrink,Dwaveflux,Dwavecoupler}. Although weakly-coupled combinations of these simple components have already allowed a large variety of applications to be explored, this is very likely a vanishingly-small subset of what is possible in the more general case of larger Josephson quantum circuits with strong nonlinearity, and couplings between degrees of freedom that are non-perturbative.

In this work, we present a new physical and numerical framework for treating such circuits. Its purpose is both to enable and to encourage the exploration of a much broader range of quantum circuit designs and modes of operation than the small set which are currently well-understood and in general use. Although this framework is built on the foundation laid by previous works on superconducting circuit quantization \cite{yurke,devoret,burkhardPRB,divincenzocircuit}, it contains two fundamentally new elements, which underpin its descriptive power: First, we use the combination of a physically-motivated canonical coordinate representation and a numerically efficient basis choice to reduce the size and increase the sparsity of the physical-level Hamiltonian matrices to be diagonalized. Second, we use a hierarchical diagonalization technique (related to tensor network ~\cite{orusTN} and renormalization group methods ~\cite{wilson,RGRMP}) in which complex quantum circuit Hamiltonians are diagonalized in multiple stages, to find approximate eigenstates in what would otherwise be intractably large Hilbert spaces. The methods we describe here have allowed us to simulate larger and more complex Josephson quantum circuits than previously possible \cite{JPSQ}.

We begin in section ~\ref{s:classH} by describing the classical circuit Hamiltonian and our chosen coordinate representation. We quantize this Hamiltonian in section ~\ref{s:quantH}, and describe the specific basis states used for each kind of canonical mode. Section ~\ref{s:JPSQ} contains a detailed example of the use of our method to describe the recently-proposed Josephson phase-slip qubit (JPSQ) circuit \cite{JPSQ}. Next, in sections ~\ref{s:partition} and ~\ref{s:truncation} we describe the first stage of the hierarchical diagonalization process, apply it to the example circuit from section ~\ref{s:JPSQ}, and compare the results to those obtained previously in that section. We conclude with a summary and future prospects in section ~\ref{s:conclusion}.

\section{Classical circuit Hamiltonian}\label{s:classH}

Following previous works on superconducting circuit quantization \cite{yurke,devoret,burkhardPRB,divincenzocircuit}, consider a general circuit of $N_\textrm{n}$ nodes $i\in\{1...N_\textrm{n}\}$ and $N_\textrm{b}$ branches $i\in\{1...N_\textrm{b}\}$, containing capacitors, linear inductors, and Josephson junctions, with the Hamiltonian:

\begin{equation}
H_\textrm{tot}=H_\textrm{LC}+U_\textrm{J}\label{eq:HCl}
\end{equation}

\noindent The first term describes the linear circuit elements, and can be written:
\begin{eqnarray}
H_\textrm{LC}&=&\frac{1}{2}\Bigl[\vec{V}_\textrm{n}^\mathrm{T}\cdot\mathbf{C}_\textrm{n}\cdot\vec{V}_\textrm{n}+\vec{I}_\textrm{b}^\mathrm{T}\cdot(\mathbf{L}_\textrm{b}+\mathbf{M})\cdot\vec{I}_\textrm{b}\Bigr]\nonumber\\
&&-\;\vec{Q}_\textrm{n}^\mathrm{T}\cdot\vec{V}_\textrm{n}^\textrm{e}-\vec{\Phi}_\textrm{b}^\mathrm{T}\cdot\vec{I}_\textrm{b}^\textrm{e}\label{eq:HLCIV}\\
&&+\;\frac{1}{2}\Bigl[(\vec{V}_\textrm{n}^\textrm{e})^\textrm{T}\cdot\mathbf{C}_\textrm{n}\cdot\vec{V}_\textrm{n}^\textrm{e}+(\vec{I}_\textrm{b}^\textrm{e})^\textrm{T}\cdot(\mathbf{L}_\textrm{b}+\mathbf{M})\cdot\vec{I}_\textrm{b}^\textrm{e}\Bigr]\hspace{0.25cm}
\nonumber
\end{eqnarray}

\noindent where the vectors $\vec{V}_\textrm{n}$ (length-$N_\textrm{n}$) represent the node voltages, the vectors $\vec{I}_\textrm{b}$ (length-$N_\textrm{b}$) represent the branch inductor currents, and $\vec{V}_\textrm{n}^\textrm{e},\vec{I}_\textrm{b}^\textrm{e}$ are the external node bias voltages and branch bias currents. The matrix $\mathbf{C}_\textrm{n}$ is the ($N_\textrm{n}\times N_\textrm{n}$) symmetric node capacitance matrix, whose diagonal elements are the sum of all capacitances connected to each node, and whose off-diagonal elements are $-1$ times the sum of all capacitances connecting each pair of nodes. The matrix $\mathbf{L}_\textrm{b}$ is the ($N_\textrm{b}\times N_\textrm{b}$) diagonal branch inductance matrix, whose diagonal elements are the self-inductances of each branch, while the $N_\textrm{b}\times N_\textrm{b}$ mutual inductance matrix $\mathbf{M}$ is zero on the diagonal and has off-diagonal elements given by the mutual inductances between each pair of branches (with a sign according to the relative alignment of the two mutually-coupled branches). The relationships between node voltages and charges, and branch currents and fluxoids are given by the constitutive relations:

\begin{eqnarray}
\vec{V}_\textrm{n}&\equiv&\mathbf{C}^{-1}_\textrm{n}\cdot\vec{Q}_\textrm{n}\label{eq:constitute0E}\\
\vec{I}_\textrm{b}&\equiv&(\mathbf{L}_\textrm{b}+\mathbf{M})^{-1}\cdot\vec{\Phi}_\textrm{b}\label{eq:constitute0}
\end{eqnarray}

The second line of eq.~\ref{eq:HLCIV} describes the interaction between the circuit and its classical bias current and voltage sources $\vec{I}_\textrm{b}^\textrm{e}$ and $\vec{V}_\textrm{b}^\textrm{e}$, respectively, and the last line is the potential energy associated with those sources. Note that we are therefore assuming at this point that the linear circuit networks associated with the bias sources are explicitly included in the circuit being analyzed. This is a departure from the usual treatments of superconducting circuits, in which external bias sources are treated by writing \textit{charge} and/or \textit{fluxoid} offsets directly into the Hamiltonian at the start. Although the latter method is simpler and can give quantitatively correct results in most cases if used correctly, it can also lead to physical misconceptions, so for pedagogical purposes we start from a strict, physical description in order to illustrate both how the simplified version arises and how and when it can be correctly used. Strictly speaking, it is unphysical to speak of a ``flux bias" or a ``charge bias" source in the present context. Rather, these two terms might loosely be said to refer to the physical limit in which the bias source is infinitely stiff (i.e., has an infinitely large source inductance or capacitance, respectively), such that the current or voltage it supplies is independent of the circuit to which it is connected.

The Josephson potential energy in eq.~\ref{eq:HCl} is given by:

\begin{equation}
U_\textrm{J}=\sum_{\alpha\in J}^{N_\textrm{J}} E_{\textrm{J}\alpha}\left[1-\cos\phi_{\textrm{J}\alpha}\right]\label{eq:UJ}
\end{equation}

\noindent where $E_{\textrm{J}\alpha}$ and $\phi_{\textrm{J}\alpha}$ are, respectively, the Josephson energy and the gauge-invariant phase difference across the $\alpha^\textrm{th}$ Josephson junction.


In order to quantize the total Hamiltonian of eqs.~\ref{eq:HCl}-\ref{eq:UJ}, we must first specify a set of canonical coordinates and conjugate momenta, and express it in terms of these variables. Based on the usual quantum network theory prescription \cite{yurke,devoret}, the first step is to define a classical spanning tree, which is the lumped-circuit equivalent of fixing a gauge. For the present purpose, and for reasons that will become clear shortly, we define a ``superconducting spanning tree" $\mathcal{S}$ as follows: Starting from the graph $\mathcal{G}$ whose edges are the circuit's branches, we construct the subgraph $\mathcal{G}_L\in\mathcal{G}$ associated with the circuit's inductive and Josephson branches. If every node in the circuit is both contained in $\mathcal{G}_L$, and has at least one path connecting it to ground which is contained in $\mathcal{G}_L$, then a superconducting spanning tree $\mathcal{S}$ is any subgraph of $\mathcal{G}_L$ with the property that from each node in the circuit $\mathcal{S}$ contains exactly one path to ground.

If $\mathcal{G}_L$ does not include at least one path to ground from every node in the circuit (i.e, there exist some part(s) of the circuit which have no DC superconducting connection to ground), the procedure must be modified. We first identify the connected components $\{\mathcal{G}_L^i\}$ of $\mathcal{G}_L$. The subset of these that do not contain ground correspond to what we refer to below as the ``islands" of the circuit. For each of these island subgraphs, we choose a single node which will act as a ``virtual ground". We then perform the procedure described in the previous paragraph to find individual spanning trees for each of the connected components $\{\mathcal{G}_L^i\}$ (using the chosen virtual ground node in place of actual ground for any $\{\mathcal{G}_L^i\}$ that do not contain ground). The union of the resulting set of spanning trees then constitutes $\mathcal{S}$ \footnote{Note that the resulting graph is not actually a single tree, but only a union of disconnected trees.}, having the generalized spanning property that from any node in the circuit, it contains exactly one path to the (virtual or real) ground node of the connected component $\mathcal{G}_L^i$ which contains that node.

The set of branches $\mathcal{C}\equiv\mathcal{G}_L-\mathcal{S}$ are the circuit's superconducting ``closure" branches, so named because each one defines a closed superconducting loop $\ell_{ij}$ containing only inductors or Josephson junctions (formed by the two paths on $\mathcal{S}$ to (virtual or real) ground from nodes $i$ and $j$). Each such loop must satisfy the fluxoid quantization condition \cite{foundations}, which requires that the (directed) branch fluxoids around the loop sum to an integer number $m_{ij}$ of fluxoid quanta $\Phi_0\equiv h/2e$:

\begin{equation}
\Phi_{b_{ij}}+\sum_{b_{pq}\in \ell_{ij},\mathcal{S}}\Phi_{b_{pq}}=m_{ij}\Phi_0, \;\;m_{ij}\in\mathbb{Z}\label{eq:fquant}
\end{equation}

\noindent Based on this, the circuit's $N_\textrm{n}$ canonical node fluxoid variables $\Phi_i$ ($i\in 1...N_\textrm{n}$) can be defined by requiring that the directed branch fluxoid $\Phi_{ij}$ associated with branch $b_{ij}$ be given by:

\begin{equation}
\Phi_{b_{ij}} = \begin{cases}
  \Phi_j-\Phi_i & b_{ij}\in\mathcal{S} \\
  \Phi_j-\Phi_i+m_{ij}\Phi_0 & b_{ij}\in\mathcal{C}\label{eq:node}
\end{cases}
\end{equation}

\noindent where the fluxoid quantum number $m_{ij}$ will not be an observable quantity for any loop interrupted by one or more Josephson junctions. The canonical momenta conjugate to the node fluxoid variables of eq.~\ref{eq:node} are the node charge variables $Q_i$, given for each node by the total charge on all capacitors connected to it.

We highlight the fact that the second relation in eq.~\ref{eq:node} is different from the usual formulation, in which a $c$-number external flux is directly specified for each superconducting loop $\ell_{ij}$, and is added explicitly to the corresponding closure branch flux [c.f., eq. 2.13 in the seminal work of ref. ~\onlinecite{devoret}]. In the formulation above, external bias fluxes are coupled to individual \textit{branch} inductances of the circuit, via $\mathbf{M}$ in eq.~\ref{eq:HLCIV}, and the $c$-number bias \textit{currents} $\vec{I}_\textrm{b}^\textrm{e}$ (themselves flowing in the branch inductors of bias networks explicitly included in the circuit description). The external bias flux offsets are then contained in the physical branch flux coordinates themselves via the constitutive relation eq.~\ref{eq:constitute0}, rather than appearing explicitly in eq.~\ref{eq:node}. This approach, although not as simple as that used in ref.~\onlinecite{devoret}, is closer to physical reality, since no ``external" flux can be applied to a loop unless it has nonzero geometric inductance. It also avoids confusion that can otherwise arise between loops in the graph-theoretic vs. the geometric sense; this is particularly important when treating circuits with loops that share inductive branches, such as the tunable flux qubit ~\cite{orlando,mooij,Dwaveflux,Dwavecoupler,MITLLflux,MITLLcoupled} and the Josephson phase-slip qubit \cite{JPSQ} discussed below.

Before the circuit can be quantized, we need to express the magnetic part of eq.~\ref{eq:HLCIV} in terms of the canonical node fluxoid coordinates $\vec{\Phi}_\textrm{n}$ using the transformation:

\begin{eqnarray}
\vec{I}_\textrm{b}=\mathbf{R}_\textrm{bn}\cdot\vec{I}_\textrm{n}\nonumber\\
\vec{\Phi}_\textrm{b}=\mathbf{R}_\textrm{bn}\cdot\vec{\Phi}_\textrm{n}
\end{eqnarray}

\noindent where the branch matrix $\mathbf{R}_\textrm{bn}$ consists of $N_\textrm{b}$ row vectors each of length-$N_\textrm{n}$, and describing one of the $N_\textrm{b}$ directed inductive branches with a +1 entry in the position of its starting node, and a -1 entry for its ending node (if it is not ground) \footnote{The branch matrix $\mathbf{R}_\textrm{bn}$ serves a similar function to, though is not the same as, the quantity $S_{nb}$ of ref.~\onlinecite{devoret}.}. This allows the magnetic constitutive relation of eq.~\ref{eq:constitute0} to be replaced with:

\begin{equation}
\vec{I}_\textrm{n}\equiv\mathbf{L}^{-1}_\textrm{n}\cdot\vec{\Phi}_\textrm{n}\label{eq:nconstitute}
\end{equation}

\noindent where we have defined the inverse node inductance matrix (which is itself not invertible if the circuit has any non-oscillator modes):

\begin{equation}
\mathbf{L}^{-1}_\textrm{n}=\mathbf{R}_\textrm{bn}^\textrm{T}\cdot(\mathbf{L}_\textrm{b}+\mathbf{M})^{-1}\cdot \mathbf{R}_\textrm{bn}
\end{equation}

\noindent and the resulting node current vector $\vec{I}_\textrm{n}$ gives the total current flowing from each node into all closure branches connected to it by the spanning tree. Equation ~\ref{eq:HLCIV} can now be written in the canonical node representation as:

\begin{eqnarray}
H_\textrm{LC}&=&\frac{1}{2}\Bigl[(\vec{Q}_\textrm{n}-\vec{\delta Q}_\textrm{n})^\mathrm{T}\cdot\mathbf{C}^{-1}_\textrm{n}\cdot(\vec{Q}_\textrm{n}-\vec{\delta Q}_\textrm{n})\nonumber\\
&&+(\vec{\Phi}_\textrm{n}-\vec{\delta\Phi}_\textrm{n})^\mathrm{T}\cdot\mathbf{L}^{-1}_\textrm{n}\cdot(\vec{\Phi}_\textrm{n}-\vec{\delta\Phi}_\textrm{n})\Bigr]\nonumber\\
\vec{\delta Q}_\textrm{n}&\equiv&\mathbf{C}_\textrm{n}\cdot\vec{V}_\textrm{n}^\mathrm{e}\hspace{3.7cm}\nonumber\\
\vec{\delta\Phi}_\textrm{n}&\equiv&\left(\mathbf{L}_\textrm{n}^{-1}\right)^+\cdot\mathbf{R}_\textrm{bn}^\textrm{T}\cdot\vec{I}_\textrm{b}^\mathrm{e}\label{eq:HLC}
\end{eqnarray}

\noindent where in the last line the superscript $+$ indicates the pseudo-inverse. Equation ~\ref{eq:HLC} is now in the form typically encountered at the start in most treatments of specific superconducting circuits, where the influence of the bias sources appears explicitly in the form of apparent offsets to the fluxoid and charge variables. We can now clarify the reasons for not starting from this form. First of all, since the node capacitance and inductance matrices include the biasing portion of the circuits being analyzed, their inverses will correctly capture the loading effects, including renormalization of circuit inductances and capacitances, as well as additional physical couplings between variables mediated by the bias network. Such effects can sometimes have unforseen observable consequences that would otherwise be excluded from the analysis. In addition, the derivation of eq.~\ref{eq:HLC} starting from eq.~\ref{eq:HLCIV} makes the nature of the apparent offsets clear and explicit. In the electric case, there are no physical ``offset charges" on any circuit nodes in the strictest sense (we note, however, that this term is frequently used in the literature), which would correspond to a departure from charge neutrality. Rather, an apparent charge offset appears in the Hamiltonian when the electrostatic potential of a node is changed (due to an externally-applied voltage on a gate electrode to which it is capacitively coupled); this corresponds to the electric polarization $\mathbf{P}$ in continuum electrostatics. For the magnetic case, the flux offset for an inductive branch arises from the additional term in the constitutive relation of eq.~\ref{eq:constitute0} due to the mutually-coupled flux (and therefore does not explicitly affect the current flowing in that branch).

The last step in expressing the Hamiltonian in terms of a set of canonical coordinates is to express eq.~\ref{eq:UJ} in the node representation. To do this, we need only use eqs.~\ref{eq:node}, since in the absence of external fluxes penetrating the junction barriers, the gauge-invariant phase across the junction on branch $b_{ij}$ is simply proportional to the branch flux: $\phi_{\textrm{J}_{ij}}\equiv 2\pi\Phi_{b_{ij}}/\Phi_0$.

We are now in a position to quantize the circuit by interpreting the canonically conjugate node fluxoid/charge (coordinate/momentum) pairs as quantum-mechanical operators obeying the canonical commutation relations. However, for reasons that will become clear shortly, we first transform from these node coordinates into a new canonical representation which is both conceptually and numerically superior. Any such transformation (which must preserve the canonical commutation relations) can be expressed in terms of a nonsingular $N_\textrm{n}\times N_\textrm{n}$ matrix $\mathcal{R}$:

\begin{eqnarray}
\vec{\Phi}&=&\mathcal{R}\cdot\vec{\Phi}_\textrm{n}\nonumber\\
\vec{Q}&=&(\mathcal{R}^T)^{-1}\cdot\vec{Q}_\textrm{n}
\label{eq:R}
\end{eqnarray}

\noindent where $\vec{\Phi}$ and $\vec{Q}$ are the circuit's fluxoid and charge vectors in the new coordinate representation, and the inverse inductance and capacitance matrices are transformed according to:

\begin{eqnarray}
\mathbf{L}^{-1}&=&(\mathcal{R}^T)^{-1}\cdot\mathbf{L}_\textrm{n}^{-1}\cdot\mathcal{R}^{-1}\nonumber\\
\mathbf{C}^{-1}&=&\mathcal{R}\cdot\mathbf{C}_\textrm{n}^{-1}\cdot\mathcal{R}^T\label{eq:R_LC}
\end{eqnarray}

\noindent We consider only members of a specific subset of all possible such transformations, which separate the circuit's canonical fluxoid coordinates into three distinct mode types:

\begin{itemize}
\item $N_\textrm{O}$ oscillator modes, whose fluxoid coordinates are spanned by the eigenvectors of $\mathbf{L}_\textrm{n}^{-1}$ with nonzero eigenvalues (corresponding to the principal curvatures of the linear inductive potential); $N_\textrm{O}$ is equal to the rank of $\mathbf{L}_\textrm{n}^{-1}$.
\item $N_\textrm{I}$ island modes, whose fluxoid coordinates do not appear in the Hamiltonian at all; $N_\textrm{I}$ is equal to the number of connected components of the graph $\mathcal{B}$.
\item $N_\textrm{J}\equiv N_\textrm{n}-N_\textrm{I}-N_\textrm{O}$ Josephson modes, of which the Hamiltonian is (nontrivially) invariant under a $\Phi_0$ translation.
\end{itemize}

\noindent Note that the number Josephson modes need not be equal to the number of Josephson junctions, and the number of oscillator modes need not be equal to the number of inductors.

Before proceeding, we also highlight a point about the island modes just defined.  Every circuit with island modes will have at least one such mode that is not coupled electrostatically to any other, and in many cases is not coupled to a bias source either. A natural example of this occurs for a circuit which is floating with respect to ground and biased purely by magnetic flux through a mutual inductance. In such a case, one might be tempted to simplify the circuit by connecting one of the nodes of such an island to ground, based on the intuition that if one need not keep track of the electrostatic energy of the island, its potential with respect to ground is arbitrary. Although this intuition is correct, insofar as the island's electrostatic energy is decoupled from the rest of the modes, grounding one of the nodes of such an island will in general produce \textit{additional} electrostatic couplings between its other internal modes, and thus observable changes in its properties.

After transforming to these coordinates using eqs.~\ref{eq:R} and ~\ref{eq:R_LC}, we can partition $\mathbf{C}^{-1}$ and $\mathbf{L}^{-1}$ into block submatrices corresponding to the three mode types, which we label using the subscripts $\{\textrm{O,I,J}\}$:

\begin{align}
&\phantom{=}\hspace{0.3cm}\xlongleftrightarrow[]{\hspace{0.25cm}N_\textrm{O}\hspace{0.2cm}}\;\xlongleftrightarrow[]{\hspace{0.15cm}N_\textrm{I}\hspace{0.15cm}}\;\xlongleftrightarrow[]{\hspace{0.25cm}N_\textrm{J}\hspace{0.2cm}}\;&
&&&\phantom{=}\hspace{0.3cm}\hspace{0.4cm}&&\nonumber\\
\mathbf{C}^{-1}&=\left(
\begin{BMAT}(@)[0pt,3cm,3.5cm]{c.c.c}{c.c.c}
\phantom{\Biggl[}\mathbf{C}^{-1}_\textrm{O}\phantom{\Biggr]} & \;\mathbf{C}^{-1}_\textrm{OI}\; & \mathbf{C}^{-1}_\textrm{OJ}\\
 & \phantom{\Bigl[}\mathbf{C}^{-1}_\textrm{I}\phantom{\Bigr]} & \mathbf{C}^{-1}_\textrm{IJ}\\
 &  & \phantom{\biggl[}\mathbf{C}^{-1}_\textrm{J}\phantom{\biggr]}
\end{BMAT}
\right)\;&\hat{\vec{Q}}&=\left(
\begin{BMAT}(@)[0pt,0.5cm,3.5cm]{c}{c.c.c}
\phantom{\Biggl[}\vec{Q}_{\textrm{O}}\phantom{\Biggr]} \\
\phantom{\Bigl[}\vec{Q}_{\textrm{I}}\phantom{\Bigr]} \\
\phantom{\biggl[}\vec{Q}_{\textrm{J}}\phantom{\biggr]}
\end{BMAT}
\right)\;\nonumber
\end{align}

\begin{align}
\mathbf{L}^{-1}&=\left(
\begin{BMAT}(@)[0pt,3cm,3.5cm]{c.cc}{c.cc}
\phantom{\Biggl[}\mathbf{L}^{-1}_\textrm{O}\phantom{\Biggr]} & \hspace{0.3cm}0\hspace{0.3cm} & \hspace{0.2cm}0\hspace{0.2cm}\\
 & \phantom{\Bigl[}\hspace{0.2cm}0\hspace{0.2cm}\phantom{\Bigr]} & \hspace{0.2cm}0\hspace{0.2cm}\\
 &  & \phantom{\biggl[}\;\hspace{0.2cm}0\hspace{0.2cm}\phantom{\biggr]}
\end{BMAT}
\right)\;&\hat{\vec{\Phi}}&=\left(
\begin{BMAT}(@)[0pt,0.5cm,3.5cm]{c}{c.c.c}
\phantom{\Biggl[}\vec{\Phi}_{\textrm{O}}\phantom{\Biggr]} \\
\phantom{\Bigl[}\vec{\Phi}_{\textrm{I}}\phantom{\Bigr]} \\
\phantom{\biggl[}\vec{\Phi}_{\textrm{J}}\phantom{\biggr]}
\end{BMAT}
\right)\;\label{eq:Cinv}
\end{align}

\noindent where the only nonzero block of $\mathbf{L}^{-1}$ is the submatrix $\mathbf{L}_\textrm{O}^{-1}$, which is now by construction invertible, since its eigenvalues correspond to the principal curvatures of the linear inductive potential. We can now rewrite eq.~\ref{eq:R_LC} as:

\begin{eqnarray}
H_\textrm{LC}&=&\frac{1}{2}\Bigl[(\vec{Q}-\vec{\Delta Q})^\mathrm{T}\cdot\mathbf{C}^{-1}\cdot(\vec{Q}-\vec{\Delta Q})\nonumber\\
&&+(\vec{\Phi}-\vec{\Delta\Phi})^\mathrm{T}\cdot\mathbf{L}^{-1}\cdot(\vec{\Phi}-\vec{\Delta\Phi})\Bigr]\label{eq:HLC2}
\end{eqnarray}

\begin{eqnarray}
\vec{\Delta Q}&\equiv&\mathcal{R}\cdot\mathbf{C}_\textrm{n}\cdot\vec{V}_\textrm{n}^\mathrm{e}
=\left(
\begin{BMAT}(@)[0pt,0.5cm,1.2cm]{c}{ccc}
\vec{\Delta Q}_\textrm{O}\\
\vec{\Delta Q}_\textrm{J} \\
\vec{\Delta Q}_\textrm{I}
\end{BMAT}
\right)\label{eq:dQvec}\\
\vec{\Delta\Phi}&\equiv&(\mathcal{R}^{-1})^\textrm{T}\cdot\mathbf{L}_\textrm{n}\cdot\mathbf{R}_\textrm{bn}^\textrm{T}\cdot\vec{I}_\textrm{b}^\mathrm{e}
=\left(
\begin{BMAT}(@)[0pt,0.5cm,1.2cm]{c}{ccc}
\vec{\Delta\Phi}_\textrm{O}\\
0 \\
0
\end{BMAT}
\right)\label{eq:dphivec}
\end{eqnarray}


\noindent where the second equality in eq.~\ref{eq:dphivec} follows from the fact that, according to the mode definitions above, external bias currents can only produce flux offsets of oscillator modes.


\section{Quantum Hamiltonian}\label{s:quantH}

We can now quantize in the usual manner, treating each classical coordinate/momentum (fluxoid/charge) pair as quantum operators obeying the canonical commutation relations:

\begin{equation}
\left[[\hat{\vec{\Phi}}]_i,[\hat{\vec{Q}}]_j\right]=\mathrm{i}\hbar\delta_{ij}\label{eq:commute}
\end{equation}

\noindent where the notation $[\hat{\vec{X}}]_i$ denotes the quantum operator for the $i^\textrm{th}$ entry in the vector $\vec{X}$. The total quantum Hamiltonian can be written:

\begin{equation}
\hat{H}_\textrm{tot}=\hat{H}_\textrm{O}+\hat{H}_\textrm{I}+\hat{H}_\textrm{J}+\hat{H}_\textrm{int}
\end{equation}

\noindent with the definitions:

\begin{align}
\hat{H}_\textrm{O}&=\frac{1}{2}\Bigl[\hat{\vec{\delta Q}}_\textrm{O}^\textrm{T}\cdot\mathbf{C}^{-1}_\textrm{O}\cdot\hat{\vec{\delta Q}}_\textrm{O}+\hat{\vec{\delta\Phi}}_\textrm{O}^\textrm{T}\cdot\mathbf{L}_\textrm{O}^{-1}\cdot\hat{\vec{\delta\Phi}}_\textrm{O}\Bigr]+\hat{U}_\textrm{J}(\hat{\vec{\Phi}}_\textrm{O})
\label{eq:HO}\\
\hat{H}_\textrm{I}&=\frac{1}{2}\hat{\vec{\delta Q}}_\textrm{I}^\textrm{T}\cdot\mathbf{C}^{-1}_\textrm{I}\cdot\hat{\vec{\delta Q}}_\textrm{I}\label{eq:HI}
\end{align}

\begin{eqnarray}
\hat{H}_\textrm{J}&=&\frac{1}{2}\hat{\vec{\delta Q}}^\textrm{T}_\textrm{J}\cdot\mathbf{C}^{-1}_\textrm{J}\cdot\hat{\vec{\delta Q}}_\textrm{J}+\hat{U}_\textrm{J}(\hat{\vec{\Phi}}_\textrm{J})\hspace{3cm}\label{eq:HJ}\\
\hat{H}_\textrm{int}&=&\hat{\vec{\delta Q}}_\textrm{O}^\textrm{T}\cdot\left[\mathbf{C}^{-1}_\textrm{OJ}\cdot\hat{\vec{\delta Q}}_\textrm{J}
+\mathbf{C}^{-1}_\textrm{OI}\cdot\hat{\vec{\delta Q}}_\textrm{I}\right]\nonumber\\
&&\hspace{1cm}+\hat{\vec{\delta Q}}_\textrm{J}^\textrm{T}\cdot\mathbf{C}^{-1}_\textrm{IJ}\cdot\hat{\vec{\delta Q}}_\textrm{I}
+\hat{U}_\textrm{J}(\hat{\vec{\Phi}}_\textrm{O},\hat{\vec{\Phi}}_\textrm{J})\label{eq:Hint}\\
\hat{\vec{\delta Q}}&\equiv&\hat{\vec{Q}}-\vec{\Delta Q}\hspace{0.5cm}
\hat{\vec{\delta\Phi}}\equiv\hat{\vec\Phi}-\vec{\Delta\Phi}
\end{eqnarray}

\noindent where eqs.~\ref{eq:HO}-\ref{eq:HJ} describe the energy of the oscillator, island, and Josephson subspaces, and eq.~\ref{eq:Hint} describes the couplings \textit{between} them. We have explicitly separated the parts of $\hat{U}_\textrm{J}$ according to their fluxoid coordinate dependence:

\begin{equation}
\hat{U}_\textrm{J}\equiv\hat{U}_\textrm{J}(\hat{\vec\Phi}_\textrm{J})+\hat{U}_\textrm{J}(\hat{\vec\Phi}_\textrm{O})+\hat{U}_\textrm{J}(\hat{\vec\Phi}_\textrm{O},\hat{\vec\Phi}_\textrm{J})\label{eq:UJparts}
\end{equation}

\noindent where dependence on $\hat{\vec\Phi}_\textrm{I}$ is absent by construction.

\subsection{Oscillator modes}\label{ss:oscmodes}

Consider eq.~\ref{eq:HO}, which describes a set of coupled, nonlinear oscillators. We can associate with the linear part of each of these a characteristic frequency and impedance:

\begin{equation}
\omega_i\equiv\sqrt{[\mathbf{C}^{-1}_\textrm{O}]_{ii}[\mathbf{L}_\textrm{O}^{-1}]_{ii}}\hspace{0.5cm}
Z_i\equiv\sqrt{\frac{[\mathbf{C}^{-1}_\textrm{O}]_{ii}}{[\mathbf{L}_\textrm{O}^{-1}]_{ii}}}
\end{equation}

\noindent where $i$ indicates the $i^\textrm{th}$ mode. The length-$N_\textrm{O}$ fluxoid and charge vectors in eq.~\ref{eq:HO} can then be written:

\begin{equation}
\left[\hat{\vec\Phi}_\textrm{O}\right]_i\equiv\sqrt{\frac{\hbar Z_i}{2}}(\hat{a}_i+\hat{a}_i^\dagger)\hspace{0.5cm}
\left[\hat{\vec{Q}}_\textrm{O}\right]_i\equiv-\textrm{i}\sqrt{\frac{\hbar}{2Z_i}}(\hat{a}_i-\hat{a}_i^\dagger)\label{eq:QL}
\end{equation}

\noindent in terms of the raising (lowering) operator $\hat{a}_i^\dagger$ ($\hat{a}_i$) for oscillator mode $i$.

The last term of eq.~\ref{eq:HO}, which describes the Josephson nonlinearity of the oscillator modes, can be written as a sum of terms, each having the form:

\begin{eqnarray}
\cos{\left[\frac{2\pi}{\Phi_0}\left(a\hat{\Phi}_{\textrm{O}i}+\Delta\Phi\right)\right]}=\nonumber\hspace{3cm}&\\
\frac{1}{2}\left[e^{\mathrm{i}\Delta\phi}\hat{\mathcal{D}}_{\textrm{O}i}(a)+e^{-\mathrm{i}\Delta\phi}\hat{\mathcal{D}}_{\textrm{O}i}(-a)\right]\hspace{0.5cm}&\label{eq:oscJ}
\end{eqnarray}

\noindent where $a$ is a $c$-number constant of order unity (arising from the transformation $\mathcal{R}$), $\Delta\Phi$ is an external flux, $\Delta\phi\equiv 2\pi\Delta\Phi/\Phi_0$, and the operator $\hat{\mathcal{D}}_{\textrm{O}i}(a)$ displaces the charge of oscillator $i$ by $2e\times a$. Similarly, the Josephson terms dependent on $\hat{\vec\Phi}_\textrm{O}$ in eq.~\ref{eq:Hint} will be of the form:

\begin{eqnarray}
\cos{\left[\frac{2\pi}{\Phi_0}\left(a\hat{\Phi}_{\textrm{O}i}+\hat{\Phi}_{\textrm{J}j}+\Delta\Phi\right)\right]}=\nonumber\hspace{3.0cm}&\\
\frac{1}{2}\left[e^{\mathrm{i}\Delta\phi}\hat{\mathcal{D}}_{\textrm{O}i}(a)\hat{\mathcal{D}}_{\textrm{J}j}^++e^{-\mathrm{i}\Delta\phi}\hat{\mathcal{D}}_{\textrm{O}i}(-a)\hat{\mathcal{D}}_{\textrm{J}j}^-\right]\hspace{0.5cm}&\label{eq:oscOJ}
\end{eqnarray}

\noindent where $\hat{\mathcal{D}}_{\textrm{J}j}^\pm$ displaces the charge of Josephson mode $j$ by $\pm 2e$ (the fact that these are the only permissible charge displacements is a direct result of the Josephson symmetry upon which the definition of these modes is based). Equations ~\ref{eq:oscJ} and ~\ref{eq:oscOJ} suggest a natural representation for the Josephson modes in terms of basis states with well-defined charge, which we discuss below in subsection ~\ref{ss:JImodes}.

In order to construct a concrete matrix representation of the $i^\textrm{th}$ oscillator mode, we truncate its Hilbert space to a maximum occupation $\nu_{\textrm{m}i}$, giving a ($\nu_{\textrm{m}i}+1)$-dimensional space. In this truncated basis, the canonical commutation relations of eq.~\ref{eq:commute} become, for oscillator modes $i$ and $j$ \cite{green,buchdahl}:

\begin{equation}
\left[[\hat{\vec{\Phi}}_\textrm{O}]_i,[\hat{\vec{Q}}_\textrm{O}]_j\right]=\mathrm{i}\hbar\delta_{ij}(1-\nu_{\textrm{m}i}|\nu_{\textrm{m}i}\rangle\langle\nu_{\textrm{m}i}|)\label{eq:commuteT}
\end{equation}

\noindent where it is understood that all physical observables should be evaluated in the large-$\nu_{\textrm{m}i}$ limit. In general, oscillator modes with Josephson nonlinearity will require larger $\nu_\textrm{m}$ (a larger basis set) to approach this limit. The corresponding operator matrices for the fluxoid and charge of the $i^\textrm{th}$ oscillator mode are:

\begin{equation}
\langle \nu|[\hat{\vec{\Phi}}_{O}]_i|\nu^\prime\rangle=\sqrt{\tfrac{\hbar Z_i}{2}}
\begingroup
\renewcommand*{\arraystretch}{1}
\begin{pmatrix}

      \;0\; & \;1\; & 0 \\

      1 & 0 & \sqrt{2} & \ddots \\

      0 & \sqrt{2} & 0 & \ddots & 0 \\

       & \ddots &  \ddots & 0 & \sqrt{\nu_{\textrm{m}i}}\\

      &  &  0 & \sqrt{\nu_{\textrm{m}i}} & 0

\end{pmatrix}\nonumber
\endgroup
\end{equation}

\begin{equation}
\langle \nu|[\hat{\vec{Q}}_{O}]_i|\nu^\prime\rangle=\textrm{i}\sqrt{\tfrac{\hbar}{2Z_i}}
\begingroup
\renewcommand*{\arraystretch}{1}
\begin{pmatrix}

      \;0\; & -1\; & 0 \\

      1 & 0 & -\sqrt{2} & \ddots \\

      0 & \sqrt{2} & 0 & \ddots & 0 \\

       & \ddots &  \ddots & 0 & -\sqrt{\nu_{\textrm{m}i}}\\

      &  &  0 & \sqrt{\nu_{\textrm{m}i}} & 0

\end{pmatrix}\nonumber
\endgroup
\end{equation}

In order to express the Josephson potential terms acting on oscillator mode fluxoid coordinates in this basis, we must evaluate the corresponding matrix elements of charge displacement operators like those shown in eq.~\ref{eq:oscJ}. These can be pre-computed for any fixed basis size, and have the form:

\begin{widetext}
\begin{equation}
\begingroup
\renewcommand*{\arraystretch}{1.5}
\langle\nu|\hat{\mathcal{D}}_{Oi}(a)|\nu^\prime\rangle
=e^{-a^2z_i/2}\begin{pmatrix}

       1 & \textrm{i}a\sqrt{z_i}& -a^2\frac{z_i}{\sqrt{2}}&\cdots\\

      \textrm{i}a\sqrt{z_i}& 1-a^2z_i & \textrm{i}a\sqrt{2z_i}\left(1-\tfrac{1}{2}a^2z_i\right)& \\

      -a^2\frac{z_\textrm{I}}{\sqrt{2}}& \textrm{i}a\sqrt{2z_i}\left(1-\tfrac{1}{2}a^2z_i\right) & 1-2a^2z_i+\tfrac{1}{2}a^4z_i^2 & \\

      \vdots & & & \ddots &\\
\end{pmatrix}\label{eq:Ldisp}
\endgroup
\end{equation}
\end{widetext}

\noindent where the dimensionless impedance of oscillator $i$ is $z_i\equiv Z_i/R_Q$ and $R_Q\equiv\Phi_0/2\textrm{e}=h/4\textrm{e}^2$ is the quantum of resistance for superconducting circuits.

We can see from eq. ~\ref{eq:Ldisp} that when these nonlinear oscillator displacements from the Josephson potential are not small, the corresponding charge displacement operators can have matrix representations that are not very sparse (i.e. have a large fraction of nonzero entries). This can make diagonalizing the matrix Hamiltonian (acting in the tensor product space of all modes) substantially more computationally intensive. Therefore, it is of particular importance to minimize the oscillator basis size in these cases. One way we can do this is to ``pre-compensate" for known displacements of the oscillator modes by external biases, whose effects would otherwise require a larger basis to capture accurately.

To do this, we can re-express the circuit Hamiltonian formally in a displaced oscillator basis using a unitary transformation:

\begin{align}
\hat{H}_\textrm{tot}|\Psi_n\rangle &= E_n|\Psi_n\rangle\nonumber\\
&\downarrow\nonumber\\
\underbrace{\left[\hat{U}\hat{H}_\textrm{tot}\hat{U}^\dagger\right]}_{\hat{H}^\prime_\textrm{tot}}\underbrace{\left[\hat{U}|\Psi_n\rangle \right]}_{|\Psi_n^\prime\rangle}&= E_n\underbrace{\left[\hat{U}|\Psi_n\rangle\right]}_{|\Psi_n^\prime\rangle}
\label{eq:basis}
\end{align}

\noindent where the transformation $\hat{U}$ is a displacement according to the bias offsets defined above in eqs.~\ref{eq:dphivec}, ~\ref{eq:dQvec}:

\begin{equation}
\hat{U}=\exp{\left[\frac{\textrm{i}}{\hbar}\left(\hat{\vec{Q}}\cdot\vec{\Delta\Phi}_\textrm{O}+\hat{\vec{\Phi}}\cdot\vec{\Delta Q}_\textrm{O}\right)\right]}\label{eq:Utrans}
\end{equation}

\noindent This has the effect of shifting the fluxoid and charge center points of each oscillator's \textit{basis set} to the points where the \textit{physical states} will be displaced under the action of the bias sources, such that:

\begin{equation}
\hat{U}\hat{\vec{\delta Q}}_\textrm{O}\hat{U}^\dagger=\hat{\vec{Q}}_\textrm{O}\hspace{1cm}
\hat{U}\hat{\vec{\delta\Phi}}_\textrm{O}\hat{U}^\dagger=\hat{\vec\Phi}_\textrm{O}\label{eq:deltabasis}
\end{equation}

\noindent We emphasize that eq.~\ref{eq:Utrans} is a basis transformation, and therefore leaves all observable quantities unaffected. Under this transformation, eqs.~\ref{eq:HO} and ~\ref{eq:Hint} become:

\begin{eqnarray}
\hat{H}_\textrm{O}^\prime&=&\frac{1}{2}\Bigl[\hat{\vec{Q}}_\textrm{O}^\textrm{T}\cdot\mathbf{C}^{-1}_\textrm{O}\cdot\hat{\vec{Q}}_\textrm{O}
+\hat{\vec{\Phi}}_\textrm{O}^\textrm{T}\cdot\mathbf{L}_\textrm{O}^{-1}\cdot\hat{\vec{\Phi}}_\textrm{O}\Bigr]\nonumber\\
&&+\hat{U}_\textrm{J}(\hat{\vec{\Phi}}_\textrm{O}+\vec{\Delta\Phi}_\textrm{O})\hspace{1cm}\label{eq:HO2}\\
\hat{H}_\textrm{int}^\prime&=&\hat{\vec{Q}}_\textrm{O}^\textrm{T}\cdot\left[\mathbf{C}^{-1}_\textrm{OJ}\cdot\hat{\vec{\delta Q}}_\textrm{J}+\mathbf{C}^{-1}_\textrm{OI}\cdot\hat{\vec{\delta Q}}_\textrm{I}\right]+\hat{\vec{\delta Q}}_\textrm{J}^\textrm{T}\cdot\mathbf{C}^{-1}_\textrm{IJ}\cdot\hat{\vec{\delta Q}}_\textrm{I}\nonumber\\
&&+\hat{U}_\textrm{J}(\hat{\vec{\Phi}}_\textrm{O}+\vec{\Delta\Phi}_\textrm{O},\hat{\vec{\Phi}}_\textrm{J})\label{eq:Hint2}\hspace{0.5cm}
\end{eqnarray}

\noindent while eqs.~\ref{eq:HJ} and ~\ref{eq:HI} remain the same. Note that there is no benefit to including the corresponding basis transformation of the Josephson and island modes, since these are only subject to charge offsets, and such offsets have only a trivial (diagonal) effect on the charge basis states (c.f., eq.~\ref{eq:chgop}).

\subsection{Josephson and island modes}\label{ss:JImodes}

For the island and Josephson modes, we use a basis of charge states $|q\rangle$ (with $q$ in Cooper pairs) such that $[\hat{\vec{\Phi}}_{J,I}]_i|q\rangle=2\mathrm{e}q|q\rangle$. We truncate the charge state basis for each such variable at maximal values $\pm q_\textrm{m}$ so that $-q_\textrm{m}\le q\le q_\textrm{m}$. In this basis, the charge operator $[\hat{\vec{Q}}_{J,I}]_i$ for mode $i$ can be written as the matrix:

\begin{equation}
\langle q|[\hat{\vec{Q}}_{J,I}]_i|q^\prime\rangle=2\mathrm{e}
\begingroup
\renewcommand*{\arraystretch}{1.25}
\begin{pmatrix}

      q_{\textrm{m}i} & 0\hspace{1.5cm}\\

      0 & q_{\textrm{m}i}-1\hspace{1.5cm} \\

       & \ddots\hspace{0.5cm}\ddots\hspace{0.5cm}\ddots &  \\

      & \hspace{1.5cm}-q_{\textrm{m}i}+1 &  0\\

      & \hspace{1.5cm}0 &  -q_{\textrm{m}i}\;\;

\end{pmatrix}\label{eq:chgop}
\endgroup
\end{equation}

Since no island mode flux variables appear in the Hamiltonian by construction, we require here only the Josephson mode flux operators. By our definition above, the classical Hamiltonian must be $\Phi_0$-periodic in each of these variables, and their quantization using a finite basis set therefore requires some care. The situation is similar to that encountered when defining an observable Hermitian operator for the phase of an electromagnetic field mode \cite{pegg}. We first construct a set of $2q_{\textrm{m}i}+1$ orthonormal ``flux basis states" $|\Phi_{ik}\rangle$ ($k\in\{-q_{\textrm{m}i}...q_{\textrm{m}i}\}$) for the $i^\textrm{th}$ Josephson mode:

\begin{eqnarray}
\langle q|\Phi_{ik}\rangle\equiv\sqrt{\frac{1}{2q_{\textrm{m}i}+1}}\exp{\left[\frac{2\pi\textrm{i} kq}{2q_{\textrm{m}i}+1}\right]}\label{eq:fluxstate}\\
\nonumber
\end{eqnarray}

\noindent defined here so that the state corresponding to zero flux ($k=0$) is composed of an equal superposition of all charge states $|q\rangle$. The flux operator for mode $i$ can then be written in terms of these states:


\begin{eqnarray}
\langle q|[\hat{\vec{\Phi}}_\textrm{J}]_i|q^\prime\rangle\equiv\Phi_0\sum_{k=-q_{\textrm{m}i}}^{q_{\textrm{m}i}}\frac{k}{2q_{\textrm{m}i}+1}\langle q|\Phi_{ik}\rangle\langle\Phi_{ik}|q^\prime\rangle\hspace{2cm}\nonumber\\
=\begin{cases}
\hspace{0.5cm}0,\;\;q=q^\prime\\
&\phantom{0}\\
\Phi_0\frac{q_{\textrm{m}i}\sin{\left[2\pi\tfrac{(q_{\textrm{m}i}+1)(q-q^\prime)}{2q_{\textrm{m}i}+1}\right]}-(q_{\textrm{m}i}+1)\sin{\left[2\pi\tfrac{q_{\textrm{m}i}(q-q^\prime)}{2q_{\textrm{m}i}+1}\right]}}{\textrm{i}(2q_{\textrm{m}i}+1)^2\left(1-\cos{\left[2\pi\tfrac{q-q^\prime}{2q_{\textrm{m}i}+1}\right]}\right)},\\
\hspace{1cm}q\neq q^\prime\label{eq:fluxop}
\end{cases}\nonumber
\end{eqnarray}

\noindent where:

\begin{eqnarray}
\langle q|[\hat{\vec{\Phi}}_\textrm{J}]_i|\Phi_{ik}\rangle=\Phi_{ik}\langle q|\Phi_{ik}\rangle\nonumber\\
\Phi_{ik}\equiv\frac{k\Phi_0}{2q_{\textrm{m}i}+1}\hspace{1cm}
\end{eqnarray}

\noindent As with the truncated oscillator basis discussed above, the canonical commutation relations of eq.~\ref{eq:commute} no longer hold for the truncated representations $\langle q|[\hat{\vec{\Phi}}_\textrm{J}]_i|q^\prime\rangle$ and $\langle q|[\hat{\vec{Q}}_\textrm{J}]_i|q^\prime\rangle$, and physically measurable quantities are understood to be evaluated in the limit of large $q_{\textrm{m}i}$.

Finally, the charge displacement operators for Josephson degrees of freedom can be represented by the matrices:

\begin{equation}
\langle q|\hat{\mathcal{D}}^+_{Ji}|q^\prime\rangle=\begin{pmatrix}

      0 \\

      1 & 0 \\

      & \ddots & \ddots \\

      & & 1 & 0

\end{pmatrix}\label{eq:DJp}
\end{equation}

\begin{equation}
\langle q|\hat{\mathcal{D}}^-_{Ji}|q^\prime\rangle=\begin{pmatrix}

       0 & 1\\

       & \ddots & \ddots \\

      & & 0 & 1 \\

      & & & 0

\end{pmatrix}\label{eq:DJm}
\end{equation}

\noindent Using the above results, the full quantum Hamiltonian can be expressed using a tensor product basis containing $\nu_\textrm{m}$ oscillator states for each oscillator mode, and $2q_\textrm{m}+1$ charge states for each island and Josephson mode \footnote{Note that eqs.~\ref{eq:Ldisp}, \ref{eq:chgop}, \ref{eq:fluxop}, and \ref{eq:DJp},\ref{eq:DJm} describe quantum \textit{operators}, which act in the Hilbert space of the $i^\textrm{th}$ circuit mode, represented as a matrix in the chosen basis. This is to be contrasted with, e.g., eq.~\ref{eq:Cinv}, whose rows and columns refer to the \textit{classical} components of a charge vector, \textit{each of which} has a quantum operator.}, with a total dimension of:

\begin{equation}
\left(\prod\limits_{i=1}^{N_\textrm{O}}(\nu_{\textrm{m}i}+1)\right)\left(\prod\limits_{i=1}^{N_\textrm{J}}(2q_{\textrm{m}i}+1)\right)\left(\prod\limits_{i=1}^{N_\textrm{I}}(2q_{\textrm{m}i}+1)\right)
\end{equation}

\section{Example: Josephson phase-slip qubit circuit}\label{s:JPSQ}

\begin{figure}[h!]
    \begin{center}
    \includegraphics[width=0.9\linewidth]{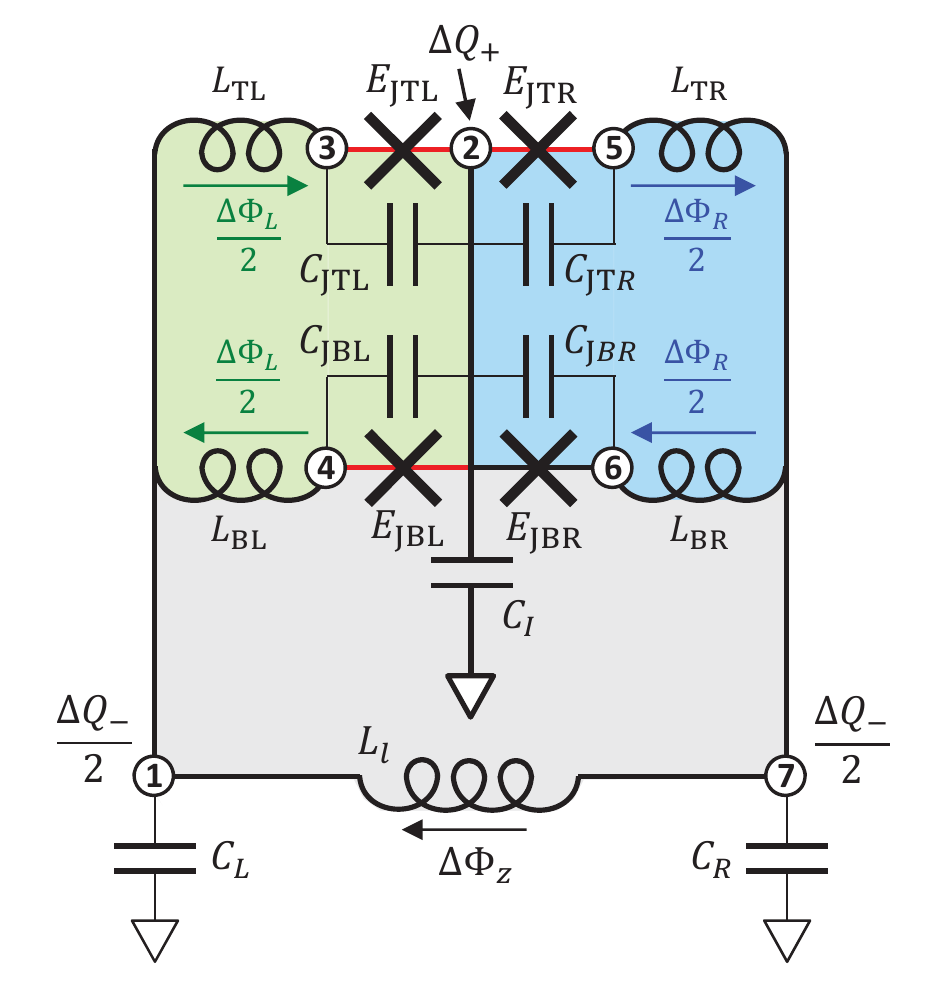}
    \caption[]
        {Example circuit: RF-SQUID-style Josephson phase-slip qubit \cite{JPSQ}. Circuit nodes are labelled with numbers, the chosen superconducting spanning tree is indicated by bold black lines, and the three superconducting closure branches are shown with bold red lines. Since the circuit is floating (i.e. it has no Josephson or inductive branches connected to ground), the spanning tree includes a capacitive branch to ground from node 2. The externally-applied fluxes $\Delta\Phi_\textrm{L}$, $\Delta\Phi_\textrm{R}$, and $\Delta\Phi_\textrm{z}$ are indicated with labelled arrows next to the inductive branches through which they are threaded, and the externally applied node offset charges $\Delta Q_\pm$ are shown next to their respective nodes 1,2, and 7.}
        \label{fig:JPSQ}
    \end{center}
\end{figure}

As a concrete example of the method just outlined, we consider the circuit of fig.~\ref{fig:JPSQ}: the RF-SQUID-style Josephson phase slip qubit from ref.~\onlinecite{JPSQ}. To keep the algebra compact, we exclude the bias circuits and specify the bias flux and charge offsets explicitly as described in the discussion of eq.~\ref{eq:HLC}. The resulting circuit has five oscillator modes, one Josephson mode, and one island mode, according to the definitions given above. We now write one possible choice for these canonical mode variables (written in terms of the node variables $\Phi_\textrm{n},Q_\textrm{n}$, where the numeric indices $n\in\{1,...,7\}$ refer to the node labels shown in the figure). For the five oscillator modes, we choose:

\begin{eqnarray}
\{\hat\Phi_{\delta},\hat{Q}_{\delta}\}&\equiv&\Biggl\{\frac{\hat\Phi_3+\hat\Phi_4}{2}-\frac{\hat\Phi_5+\hat\Phi_6}{2},\nonumber\\
&&\hspace{1cm}\frac{\hat{Q}_1+\hat{Q}_3+\hat{Q}_4}{2}-\frac{\hat{Q}_5+\hat{Q}_6+\hat{Q}_7}{2}\Biggr\}\nonumber\\
\{\hat\Phi_{\ell},\hat{Q}_{\ell}\}&\equiv&\Biggl\{\hat\Phi_7-\hat\Phi_1+\frac{\hat\Phi_3+\hat\Phi_4-\hat\Phi_5-\hat\Phi_6}{2},\frac{\hat{Q}_7-\hat{Q}_1}{2}\Biggr\}\nonumber\\
\{\hat\Phi_\textrm{p},\hat{Q}_\textrm{p}\}&\equiv&\Biggl\{\frac{\hat\Phi_7+\hat\Phi_1}{2}-\frac{\hat\Phi_3+\hat\Phi_4+\hat\Phi_5+\hat\Phi_6}{4},\hat{Q}_1+\hat{Q}_7\Biggr\}\nonumber\\
\{\hat\Phi_{\textrm{R}},\hat{Q}_{\textrm{R}}\}&\equiv&\left\{\hat\Phi_6-\hat\Phi_5,\frac{\hat{Q}_6-\hat{Q}_5}{2}\right\}\nonumber\\
\{\hat\Phi_{\textrm{L}},\hat{Q}_{\textrm{L}}\}&\equiv&\left\{\hat\Phi_3-\hat\Phi_4,\frac{\hat{Q}_3-\hat{Q}_4}{2}\right\}\label{eq:oscs}
\end{eqnarray}

\noindent The first two of these, subscripted $\delta$ and $\ell$, are loop oscillator modes (in the sense that they could be said to describe currents circulating around the loops of the qubit), and are chosen to have this particular form so that (as we will see below) only one of them ($\delta$) couples directly to the Josephson potential, while the other ($\ell$) only acquires a weak nonlinearity indirectly through linear electromagnetic interactions with other, nonlinear modes. For the single Josephson and island modes, we use:

\begin{eqnarray}
\{\hat\Phi_\textrm{J},\hat{Q}_\textrm{J}\}&\equiv&\left\{\Phi_2-\frac{\hat\Phi_3+\hat\Phi_4+\hat\Phi_5+\hat\Phi_6}{4},\hat{Q}_2-\hat{Q}_\textrm{I}\right\}\hspace{0.5cm}\label{eq:J}\\
\{\hat\Phi_\textrm{I},\hat{Q}_\textrm{I}\}&\equiv&\left\{\hat\Phi_2,\sum_{n=1}^7\hat{Q}_n\right\}\label{eq:isl}
\end{eqnarray}

The quantum Hamiltonian, written in terms of these coordinates, and after the basis transformation of eqs.~\ref{eq:basis}-\ref{eq:deltabasis}, is given by:

\begin{widetext}
\begin{eqnarray}
\hat{H}_\textrm{LC}^\prime=\sum_\alpha\hat{H}_{\textrm{O}\alpha}+\frac{(\hat{Q}_\textrm{I}+\Delta Q_-+\Delta Q_+)^2}{2C_\textrm{I}}+\frac{(\hat{Q}_\textrm{J}-\Delta Q_-)^2}{8C_{\textrm{J}+}}-
\frac{(\hat{Q}_\textrm{I}+\Delta Q_-+\Delta Q_+)\hat{Q}_\textrm{p}}{C_\textrm{I}}+\frac{(\hat{Q}_\textrm{J}-\Delta Q_-)\hat{Q}_\textrm{p}+4\hat{Q}_{\ell}\hat{Q}_{\delta}}{4C_{\textrm{J}+}}\hspace{1.0cm}
\nonumber\\
+\frac{\hat{Q}_\textrm{p}\hat{Q}_{\ell}}{2}\left[\frac{1}{C_\textrm{L}}-\frac{1}{C_\textrm{R}}\right]+\left(\hat{Q}_\textrm{J}+\hat{Q}_\textrm{p}-\Delta Q_-\right)\left[\frac{\hat{Q}_\textrm{R}}{C_{\textrm{JR}-}}-\frac{\hat{Q}_\textrm{L}}{C_{\textrm{JL}-}}\right]
+2\left(\hat{Q}_{\ell}+\hat{Q}_{\delta}\right)\left[\frac{\hat{Q}_\textrm{R}}{C_{\textrm{JR}-}}+\frac{\hat{Q}_\textrm{L}}{C_{\textrm{JL}-}}-\frac{\hat{Q}_\textrm{J}+\hat{Q}_\textrm{p}-\Delta Q_-}{4C_{\textrm{J}}-}\right]\nonumber\\
-\frac{\hat\Phi_{\ell}\hat\Phi_{\delta}}{L_\ell}+\frac{\hat\Phi_{\ell}\hat\Phi_\textrm{p}}{2}\left[\frac{1}{L_{\textrm{R}+}}-\frac{1}{L_{\textrm{L}+}}\right]+\frac{\hat\Phi_{\ell}}{4}\left[\frac{\hat\Phi_\textrm{L}}{4L_{\textrm{L}-}}+\frac{\hat\Phi_\textrm{R}}{4L_{\textrm{R}-}}\right]+
\frac{\hat\Phi_\textrm{p}}{2}\left[\frac{\hat\Phi_\textrm{R}}{4L_{\textrm{R}-}}-\frac{\hat\Phi_\textrm{L}}{4L_{\textrm{L}-}}\right]
\hspace{1cm}
\label{eq:JPSQH}
\end{eqnarray}
\end{widetext}

\noindent where we have defined the inverse capacitances and inductances:

\begin{eqnarray}
\frac{1}{C_{\textrm{JR}\pm}}&=&\frac{1}{4C_\textrm{JTR}}\pm\frac{1}{4C_\textrm{JBR}}\nonumber\\
\frac{1}{C_{\textrm{JL}\pm}}&=&\frac{1}{4C_\textrm{JTL}}\pm\frac{1}{4C_\textrm{JBL}}\nonumber\\
\frac{1}{C_{\textrm{J}\pm}}&=&\frac{1}{C_{\textrm{JL}+}}\pm\frac{1}{C_{\textrm{JR}+}}\label{eq:Cdef}
\end{eqnarray}

\begin{eqnarray}
\frac{1}{L_{\textrm{R}\pm}}&=&\frac{1}{L_\textrm{LTR}}\pm\frac{1}{L_\textrm{LBR}}\nonumber\\
\frac{1}{L_{\textrm{L}\pm}}&=&\frac{1}{L_\textrm{LTL}}\pm\frac{1}{L_\textrm{LBL}}
\label{eq:Ldef}
\end{eqnarray}

\noindent The quantities $\hat{H}_{\textrm{O}\alpha}$ are the Hamiltonians for the five bare harmonic oscillator modes ($\alpha\in\{\delta,\ell,\textrm{p},\textrm{R},\textrm{L}\}$), which have frequencies and impedances determined by the following effective oscillator inductances and capacitances:

\begin{widetext}
\begin{align}
\frac{1}{C_\ell^\textrm{e}}&\equiv\frac{1}{C_{\textrm{J}+}}+\frac{1}{C_{\textrm{LR}+}},&\frac{1}{C_\textrm{p}^\textrm{e}}&\equiv\frac{1}{C_\textrm{I}}+\frac{1}{4C_{\textrm{J}+}}+\frac{1}{4C_{\textrm{LR}+}},&
\frac{1}{C_{\delta}^\textrm{e}}&\equiv\frac{1}{C_{\textrm{J}+}},&\frac{1}{C_\textrm{R}^\textrm{e}}&\equiv\frac{4}{C_{\textrm{JR}+}},&\frac{1}{C_\textrm{L}^\textrm{e}}&\equiv\frac{4}{C_{\textrm{JL}+}}\nonumber\\
\frac{1}{L_\ell^\textrm{e}}&\equiv\frac{1}{L_\ell}+\frac{1}{4L_{\textrm{L}+}}+\frac{1}{4L_{\textrm{R}+}},&\frac{1}{L_\textrm{p}^\textrm{e}}&\equiv\frac{1}{L_\ell}+\frac{1}{L_{\textrm{L}+}}+\frac{1}{L_{\textrm{R}+}},&
\frac{1}{L_{\delta}^\textrm{e}}&\equiv\frac{1}{L_\ell},&\frac{1}{L_\textrm{R}^\textrm{e}}&\equiv\frac{1}{4L_{\textrm{R}+}},&\frac{1}{L_\textrm{L}^\textrm{e}}&\equiv\frac{1}{4L_{\textrm{L}+}}
\end{align}
\end{widetext}

\noindent Finally, the Josephson potential is given by:

\begin{widetext}
\begin{eqnarray}
\hat{U}_\textrm{J}^\prime&=&E_\textrm{JTL}\left[1-\cos\left(\hat\phi_2-\hat\phi_3+\Delta\phi_\textrm{z}+\tfrac{\Delta\phi_\textrm{L}-\Delta\phi_\textrm{R}}{2}\right)\right]
+E_\textrm{JTR}\left[1-\cos\left(\hat\phi_5-\hat\phi_2-\Delta\phi_\textrm{R}\right)\right]\nonumber\\
&&+E_\textrm{JBL}\left[1-\cos\left(\hat\phi_2-\hat\phi_4+\Delta\phi_\textrm{z}-\tfrac{\Delta\phi_\textrm{L}+\Delta\phi_\textrm{R}}{2}\right)\right]
+E_\textrm{JBR}\left[1-\cos\left(\hat\phi_2-\hat\phi_6\right)\right]\label{eq:HJJPSQ0}\\
&\Rightarrow&-\frac{E_\textrm{JTL}}{2}\biggl[e^{\textrm{i}\left(\Delta\phi_\textrm{z}+\tfrac{\Delta\phi_\textrm{L}}{2}-\tfrac{\Delta\phi_\textrm{R}}{2}\right)}\mathcal{D}_{\textrm{J}+}\mathcal{D}_\textrm{L}\left(-\tfrac{1}{2}\right)\mathcal{D}_{\delta}\left(-\tfrac{1}{2}\right)
+\textrm{h.c.}\biggr]-\frac{E_\textrm{JTR}}{2}\biggl[e^{-\textrm{i}\Delta\phi_\textrm{R}}\mathcal{D}_{\textrm{J}+}\mathcal{D}_\textrm{R}\left(\tfrac{1}{2}\right)\mathcal{D}_{\delta}\left(\tfrac{1}{2}\right)
+\textrm{h.c.}\biggr]\nonumber\\
&&-\frac{E_\textrm{JBL}}{2}\biggl[e^{\textrm{i}\left(\Delta\phi_\textrm{z}-\tfrac{\Delta\phi_\textrm{L}}{2}-\tfrac{\Delta\phi_\textrm{R}}{2}\right)}\mathcal{D}_{\textrm{J}+}\mathcal{D}_\textrm{L}\left(\tfrac{1}{2}\right)\mathcal{D}_{\delta}\left(-\tfrac{1}{2}\right)
+\textrm{h.c.}\biggr]-\frac{E_\textrm{JBR}}{2}\biggl[\mathcal{D}_{\textrm{J}+}\mathcal{D}_\textrm{R}\left(-\tfrac{1}{2}\right)\mathcal{D}_{\delta}\left(\tfrac{1}{2}\right)
+\textrm{h.c.}\biggr]
\hspace{0.5cm}\label{eq:HJJPSQ}
\end{eqnarray}
\end{widetext}

\begin{figure*}
   \begin{center}
  \includegraphics[width=1.0\linewidth]{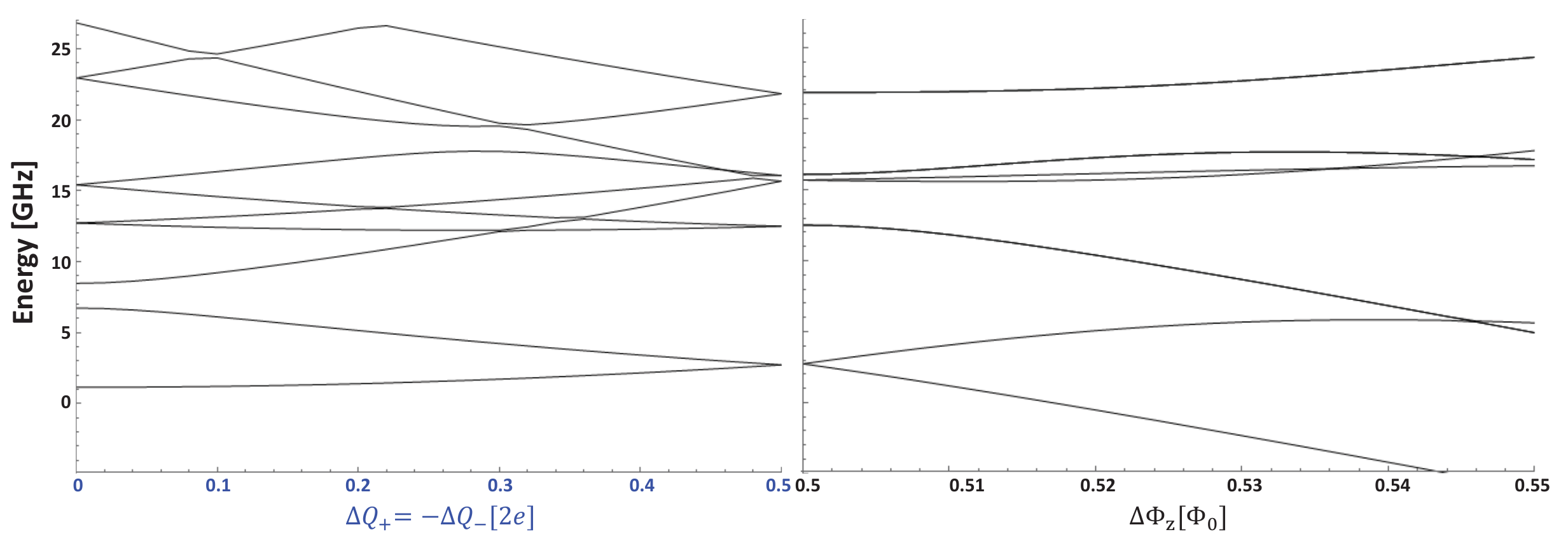}
    \caption[]
        {Calculated energy levels for the circuit of fig.~\ref{fig:JPSQ}, as a function of the island charge (left) and loop flux (right).}
        \label{fig:JPSQplot}
    \end{center}
\end{figure*}

\noindent Equation ~\ref{eq:HJJPSQ0} is written in the node phase basis, as an explicit example of how the offset phases $\Delta\phi_\textrm{z},\Delta\phi_\textrm{R},\Delta\phi_\textrm{L}$ (corresponding to the three external flux biases) contribute to the gauge-invariant phase across the three junctions which are closure branches: that is, each offset flux contributes with a sign according to whether its assumed direction (as indicated by the black, red, and blue arrows in fig.~\ref{fig:JPSQ}) is along or opposed to a path around the irreduceable loop generated by following the spanning tree from one end of the closure branch to the other. For example, for the closure branch connecting nodes 2 and 3, this path is: $2\rightarrow6\rightarrow7\rightarrow1\rightarrow3$, such that $\delta\phi_\textrm{R}/2$ has a minus sign, while $\delta\phi_\textrm{z}$ and $\delta\phi_\textrm{L}/2$ have plus signs. Equation ~\ref{eq:HJJPSQ} shows the resulting expression (neglecting a constant energy offset) when eq.~\ref{eq:HJJPSQ0} is transformed to the coordinates of eqs.~\ref{eq:oscs}-~\ref{eq:isl}, and written in terms of charge displacement operators [c.f., eq.~\ref{eq:oscOJ}]. Note that the Josephson terms do not depend on the $\ell$ oscillator mode, a result of the coordinate choice above in eq.~\ref{eq:oscs}.

Figure ~\ref{fig:JPSQplot} shows the lowest 10 energy levels calculated for the circuit of fig.~\ref{fig:JPSQ}. The truncation of the Hilbert space used for these results is: $q^\textrm{Jm}=5,q^\textrm{Im}=2$,$\nu^\textrm{Om}_\ell=3,\nu^\textrm{Om}_\textrm{p}=3,\nu^\textrm{Om}_{\delta}=4,\nu^\textrm{Om}_\textrm{R}=2,\nu^\textrm{Om}_\textrm{L}=2$, giving a total Hilbert space of $(2q^\textrm{Jm}+1)\times(2q^\textrm{Im}+1)\times(\nu^\textrm{Om}_\ell+1)\times(\nu^\textrm{Om}_\textrm{p}+1)\times(\nu^\textrm{Om}_{\delta}+1)\times(\nu^\textrm{Om}_\textrm{R}+1)\times(\nu^\textrm{Om}_\textrm{L}+1)=39,600$. This basis size is the minimum necessary for the energy splittings between the levels shown in the figure to have converged to better than $\sim1$ MHz. In general, for Josephson modes, the number of charge states required for a given level of convergence will scale as $\propto1/\sqrt{z_\textrm{J}}\propto(E_J/E_C)^{1/4}\propto(\delta\Phi_\textrm{J}/\Phi_0)^{-1}$, where $z_\textrm{J}\equiv Z_\textrm{J}/R_\textrm{Q}$ is the dimensionless impedance of the corresponding Josephson mode, $E_\textrm{C}$ is its charging energy, and $\delta\Phi_\textrm{J}$ is its ground state single-well flux uncertainty. The number of states required for oscillator modes is more dependent on the specific details of the circuit in question, though with some general trends. First, nonzero static voltages or currents require a larger basis set to describe the corresponding oscillator mode displacements; for example, persistent currents in flux qubits correspond to a flux displacement of a loop oscillator mode. Second, oscillator modes with strong Josephson nonlinearity, such as that found in the RF SQUID flux qubit, also require a larger basis set (see also, for example, ref.~\cite{wellstood}). Finally, more basis states tend to be required as the oscillator mode frequencies approach the energy range over which high precision is desired. The required number of charge states for island modes also depends on specific circuit details; however, since these modes feel no inductive potential energy, they tend to become important only when excited island charge states undergo level crossings with low-energy states of interest or when such excited charge states mediate interactions between lower-energy states of interest.

This physically-motivated representation for superconducting Josephson quantum circuits is highly compact, facilitating accurate simulation of more complex circuits than have been treated previously. Describing the inductors in the circuit by expressing them in terms of oscillator-like modes allows a basis of simple harmonic oscillator states to be used, which in nearly all cases is many times more compact than a conventional discrete variable representation in the phase basis; this is directly traceable to the fact that in contrast to the phase basis, such oscillator eigenstates already provide a natural energetic ordering, so that in most cases all of the information needed to accurately describe the low-energy properties of a circuit is already contained in a few low-energy basis states for each oscillator mode. Of course, this is mostly true because the oscillator frequencies for these modes tend to be much higher than the low-energy range of interest for the whole circuit. As the inductances and/or capacitances in a circuit are increased, this will break down, and a larger basis set will be needed. An additional source of compactness in many cases is the isolation of pure charging (island) degrees of freedom from Josephson modes. For the circuit of fig.~\ref{fig:JPSQ}, the Josephson mode required 11 charge states for $\sim$MHz convergence of the low-lying eigenvalues, while the island mode only required 3 (note that this difference could become less dramatic for parameters with less symmetry).

\section{Physical-level subsystem partitioning}\label{s:partition}

No matter how efficient the representation, the total Hilbert space required still grows exponentially with the size of the circuit. For example, full simulation of \textit{two} coupled JPSQs like that shown in fig.~\ref{fig:JPSQ}, in the manner described above, would require diagonalizing a matrix of dimension $(39,600)^2\sim10^9$, already far beyond the capabilities of any desktop computer, even for relatively sparse matrices. This would seem to pose a daunting challenge if we hope to explore the vast parameter space of superconducting quantum circuits (far beyond those simple qubits that are already well-understood), if not for two important physical facts about the range of that parameter space which is of interest to us.

First, although the Hilbert space quickly becomes enormously large for any nontrivial circuit, the actual volume of that space that we ultimately care about is usually negligibly small in comparison. Take the simple case of circuit intended for use as a qubit: although it may require tens of thousands of physical-level basis states to write the Hamiltonian down in a form we can connect to first principles, our goal is ultimately to use only the lowest few of these levels. To put it another way, the energy range of ultimate interest to us is vastly smaller than the numerical range of the Hamiltonian matrix.

Second, in most circuits of interest, there exist natural separations of energy scales. Certain degrees of freedom tend to have lower energies by design (the ones we want to use), while other degrees of freedom that we would rather ignore (but often cannot, at least not completely) or treat as bystanders, tend to have much larger energy scales, lending themselves to a perturbative treatment.

These are similar ideas to those which underly many of the numerical and analytic methods used in both quantum information (for example, tensor networks and related techniques \cite{orusTN}) and in condensed matter physics and physical chemistry (such as renormalization group methods \cite{wilson,RGRMP}): that is, an immeasurably small part of the total physical Hilbert space actually participates in the low-energy states of the system of most interest to us, so a physically-motivated representation which exploits this can be vastly more efficient than a brute force approach.

We begin by separating the terms of the total Hamiltonian of eq.~\ref{eq:HCl} to write it in the form:

\begin{eqnarray}
\hat{H}_\textrm{tot}\equiv\hat{H}_0&+&\hat{H}_1\nonumber\\
\hat{H}_0\equiv\sum_\beta^{\phantom{1}}\hat{H}_\beta,&\;&\hat{H}_1\equiv\sum_{\beta<\gamma}\hat{H}_{\beta\gamma}\label{eq:subsys}
\end{eqnarray}

\noindent where $\hat{H}_\textrm{tot}$ contains the Hamiltonians of the ``isolated" subsystems \footnote{Note, however, that these subsystem Hamiltonians already include an effective interaction, since the full capacitance and inductance matrices were inverted before the partitioning into subsystems. This is equivalent to a resummation of a subset of terms in a perturbative expansion of the interaction.} (labelled with greek letters), and $\hat{H}_\textrm{int}$ the interactions between them \footnote{The bilinear form of the electric and magnetic energies necessarily produces only pairwise interactions between subsystems; We assume here for simplicity a subsystem partitioning (and choice of canonical circuit coordinates [c.f., eq.~\ref{eq:R}]) such that the same is true for the Josephson potential, though this need not be the case in general.}. We now formally diagonalize the subsystem Hamiltonians to obtain their eigenvalues and eigenvectors in the absence of $\hat{H}_1$:

\begin{equation}
\hat{H}_\beta|k_\beta\rangle=E^\beta_k|k_\beta\rangle\label{eq:subeig}
\end{equation}

\noindent We can then construct a finite tensor product basis using the first $N_{\textrm{m}\beta}$ levels of each system $\beta$ (resulting in a total dimension $\prod_\beta N_{\textrm{m}\beta}$) and use it to re-express the Hamiltonian as:

\begin{eqnarray}
\hat{H}_\beta&=&\sum_{k}E^\beta_k|k_\beta\rangle\langle k_\beta|\otimes\left[\bigotimes_{\delta\neq\beta}\hat{I}_{\delta}\right]\label{eq:Hs}\\
\hat{H}_{\beta\gamma}&=&\sum_{k,k^\prime,l,l^\prime}V^{\beta\gamma}_{kk^\prime ll^\prime}|k_\beta\rangle\langle k^\prime_\beta|\otimes|l_\gamma\rangle\langle l^\prime_\gamma|\otimes\left[\bigotimes_{\delta\neq\beta,\gamma}\hat{I}_{s^\prime}\right]\hspace{0.75cm}\label{eq:Hs12}
\end{eqnarray}

\noindent where the $\beta-\gamma$ interaction matrix elements are given by:

\begin{eqnarray}
V^{\beta\gamma}_{kk^\prime ll^\prime}&\equiv&\vec{Q}^\beta_{kk^\prime}\cdot\mathbf{C}_{\beta\gamma}^{-1}\cdot\vec{Q}^\gamma_{ll^\prime}+\vec{\Phi}^\beta_{kk^\prime}\cdot\mathbf{L}_{\beta\gamma}^{-1}\cdot\vec{\Phi}^\gamma_{ll^\prime}\nonumber\\
&+&\biggl[\langle k_\beta|\otimes\langle l_\gamma|\biggr]\hat{U}_\textrm{J}\biggl[|k^\prime_\beta\rangle\otimes|l^\prime_\gamma\rangle\biggr]\label{eq:Veq}
\end{eqnarray}

\noindent and we have defined the subsystem charge and flux operator matrices as \footnote{Note that the second line of eq.~\ref{eq:Veq} describes an inter-subsystem interaction due to the Josephson potential, which only occurs if a junction's branch flux contains contributions from the variables of more than one of the chosen subsystems. The resulting interaction terms are of the form of eq.~\ref{eq:oscOJ}, and can be factored into subsystem operators in a similar manner to eqs.~\ref{eq:Hs12}-~\ref{eq:subops} (though in this case the interactions can involve more than two subsystems).}:

\begin{eqnarray}
\vec{Q}^\beta_{kk^\prime}&\equiv&\langle k_\beta|\hat{\vec{Q}}^\beta|k^\prime_\beta\rangle\nonumber\\
\vec{\Phi}^\beta_{kk^\prime}&\equiv&\langle k_\beta|\hat{\vec{\Phi}}^\beta|k^\prime_\beta\rangle\label{eq:subops}
\end{eqnarray}

\noindent Note that the dot products in eq.~\ref{eq:Veq} act in the \textit{classical} space of the canonical circuit coordinates and momenta (indicated by the vector notation), while the subsystem Hilbert spaces are described by the $k,k^\prime$ and $l,l^\prime$ subscripts. The coupling between these classical coordinates and momenta in different subsystems result from off-diagonal \textit{blocks} of $\mathbf{C}^{-1}$ and $\mathbf{L}^{-1}$. For example, the two-subsystem case would be written as:

\begin{eqnarray}
\mathbf{C}^{-1}&=&\left(
\begin{BMAT}(b)[0pt,2cm,1.5cm]{c:c}{c:c}
\mathbf{C}^{-1}_\beta & \mathbf{C}^{-1}_{\beta\gamma} \\
 & \mathbf{C}^{-1}_\gamma
\end{BMAT}
\right)\nonumber\\
&\phantom{=}&\;\;\;\xlongleftrightarrow[]{\hspace{0.67cm}\textrm{subsystem 1}\hspace{0.67cm}}\;\xlongleftrightarrow[]{\hspace{0.67cm}\textrm{subsystem 2}\hspace{0.67cm}}\nonumber\\
&=&\left(
\begin{BMAT}(@)[0pt,6.5cm,4cm]{c.c.c:c.c.c}{c.c.c:c.c.c}
\;\mathbf{C}^{-1}_{\textrm{O}_\beta}\; & \mathbf{C}^{-1}_{\textrm{O}_\beta\textrm{I}_\beta} & \mathbf{C}^{-1}_{\textrm{O}_\beta\textrm{J}_\beta} & \mathbf{C}^{-1}_{\textrm{O}_\beta\textrm{O}_\gamma} & \mathbf{C}^{-1}_{\textrm{O}_\beta\textrm{I}_\gamma} & \mathbf{C}^{-1}_{\textrm{O}_\beta\textrm{J}_\gamma}\\
 & \mathbf{C}^{-1}_{\textrm{I}_\beta} & \mathbf{C}^{-1}_{\textrm{I}_\beta\textrm{J}_\beta} & \;\mathbf{C}^{-1}_{\textrm{I}_\beta\textrm{O}_\gamma} & \mathbf{C}^{-1}_{\textrm{I}_\beta\textrm{I}_\gamma} & \mathbf{C}^{-1}_{\textrm{I}_\beta\textrm{J}_\gamma}\\
 &  & \mathbf{C}^{-1}_{\textrm{J}_\beta} & \;\mathbf{C}^{-1}_{\textrm{J}_\beta\textrm{O}_\gamma} & \mathbf{C}^{-1}_{\textrm{J}_\beta\textrm{I}_\gamma} & \mathbf{C}^{-1}_{\textrm{J}_\beta\textrm{J}_\gamma}\\
 &  &  & \mathbf{C}^{-1}_{\textrm{O}_\gamma} & \mathbf{C}^{-1}_{\textrm{O}_\gamma\textrm{J}_\gamma} & \mathbf{C}^{-1}_{\textrm{O}_\gamma}\\
 &  &  &  & \mathbf{C}^{-1}_{\textrm{O}_\gamma} & \mathbf{C}^{-1}_{\textrm{I}_\gamma\textrm{J}_\gamma}\\
 &  &  &  &  & \mathbf{C}^{-1}_{\textrm{J}_\gamma}
\end{BMAT}
\right)\nonumber\\
\label{eq:Cinvsub}
\end{eqnarray}

\begin{eqnarray}
\mathbf{L}^{-1}&=&\left(
\begin{BMAT}(b)[0pt,2cm,1.5cm]{c:c}{c:c}
\mathbf{L}^{-1}_\beta & \mathbf{L}^{-1}_{\beta\gamma} \\
 & \mathbf{L}^{-1}_\gamma
\end{BMAT}
\right)\nonumber\\
&\phantom{=}&\;\;\;\xlongleftrightarrow[]{\hspace{0.67cm}\textrm{subsystem 1}\hspace{0.67cm}}\;\xlongleftrightarrow[]{\hspace{0.67cm}\textrm{subsystem 2}\hspace{0.67cm}}\nonumber\\
&=&\left(
\begin{BMAT}(@)[0pt,6.5cm,4cm]{c.cc:c.cc}{c.cc:c.cc}
\mathbf{L}^{-1}_{\textrm{O}_\beta} & \phantom{\hspace{0.45cm}}0\phantom{\hspace{0.45cm}} & \phantom{\hspace{0.45cm}}0\phantom{\hspace{0.45cm}} & \;\mathbf{L}^{-1}_{\textrm{O}_\beta\textrm{O}_\gamma} & \phantom{\hspace{0.45cm}}0\phantom{\hspace{0.45cm}} & \phantom{\hspace{0.45cm}}0\phantom{\hspace{0.45cm}}\\
 & 0 & 0 & 0 & 0 & 0\\
 &  & 0 & 0 & 0 & 0\\
 &  &  & \mathbf{L}^{-1}_{\textrm{O}_\gamma} & 0 & 0\\
 &  &  &  & 0 & 0\\
 &  &  &  &  & 0
\end{BMAT}
\right)\nonumber\\
\label{eq:Linvsub}
\end{eqnarray}

\noindent The dashed lines represent the division between the two subsystems, while the dotted lines represent (as above in eqs.~\ref{eq:Cinv}) divisions between types of variables ($\textrm{O,I,J}$) within each subsystem. The subsystem blocks on the diagonal (upper left and lower right quadrants, separated by dashed lines) are associated with the internal subsystem Hamiltonians, while the off-diagonal subsystem blocks (upper right and lower left blocks, separated by dashed lines) correspond to inter-subsystem couplings.

\begin{figure}[h!]
    \begin{center}
    \includegraphics[width=0.9\linewidth]{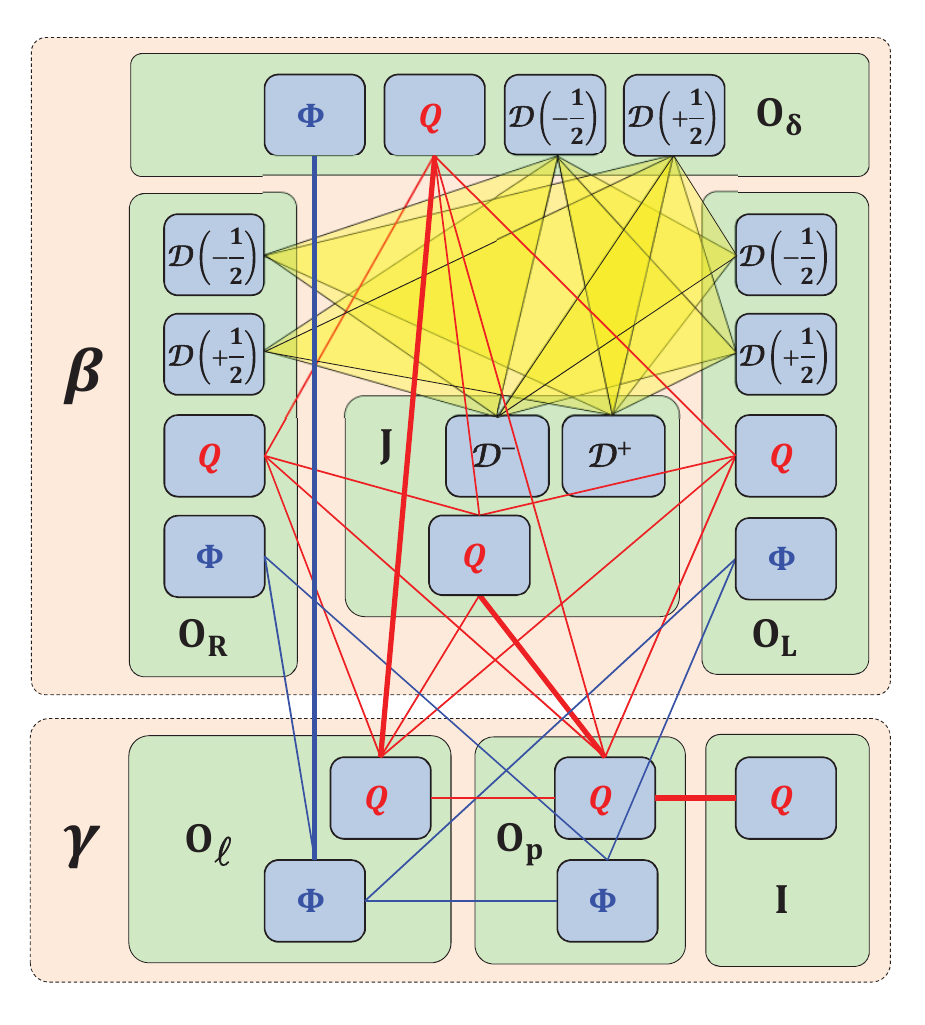}
    \caption[]
        {Mode variable interaction graph for the circuit of fig.~\ref{fig:JPSQ}. The shaded green regions indicate the quantum modes of the circuit, and the blue boxes inside them show the operators acting on each of these modes. Red lines connect pairs of charge operators and correspond to bilinear electrostatic terms, while blue lines connect pairs of flux operators, and correspond to bilinear inductive terms. Thick lines show the terms that are nonzero in the purely symmetric case, while thin lines indicate terms that are nonzero only in the presence of parameter asymmetries. Yellow triangles connect three charge displacement operators, and correspond to the Josephson potential terms of eq.~\ref{eq:HJJPSQ}. Finally, the two larger shaded pink regions indicate the two subsystems into which the Hamiltonian is partitioned in section \ref{s:partition}.}
        \label{fig:JPSQ_partition}
    \end{center}
\end{figure}

The result of this partitioning procedure is that instead of diagonalizing one very large and relatively sparse matrix, we instead diagonalize two or more much smaller sparse matrices, followed by one or more even smaller, non-sparse matrices (in general, the interaction Hamiltonian of eq.~\ref{eq:Hs12} is not sparse). To see this, we again use as an example the circuit of fig.~\ref{fig:JPSQ}. This time, however, we partition it into two subsystems $\beta,\gamma$, as illustrated in fig.~\ref{fig:JPSQ_partition}. In this figure, each quantum degree of freedom for the circuit is represented by a shaded green region, and its physical operators indicated with blue shading. Lines connecting the operators correspond to coupling terms between them in the Hamiltonian (eq.~\ref{eq:JPSQH}). For simplicity, we consider the fully symmetric case (where only the first line of eq.~\ref{eq:JPSQH} is nonzero), such that the subsystem and coupling Hamiltonians become (with $\Delta Q_+=-\Delta Q_-\equiv\Delta Q$ for simplicity):

\begin{eqnarray}
\hat{H}_\beta&=&\frac{(\hat{Q}_\textrm{J}-\Delta Q)^2}{8C_{\textrm{J}+}}+\hat{H}_{\delta}+\hat{H}_\textrm{R}+\hat{H}_\textrm{L}+\hat{H}_\textrm{J}\nonumber\\
\hat{H}_\gamma&=&\frac{(\hat{Q}_\textrm{I})^2}{2C_\textrm{I}}+\hat{H}_{\ell}+\hat{H}_\textrm{p}-\frac{\hat{Q}_\textrm{I}\hat{Q}_\textrm{p}}{C_\textrm{I}}\nonumber\\
\hat{H}_{\beta\gamma}&=&\frac{(\hat{Q}_\textrm{J}-\Delta Q)\hat{Q}_\textrm{p}+4\hat{Q}_{\ell}\hat{Q}_{\delta}}{4C_{\textrm{J}+}}-\frac{\hat\Phi_{\ell}\hat\Phi_{\delta}}{L_\ell}
\label{eq:JPSQHsub}
\end{eqnarray}

\begin{figure}[h!]
    \begin{center}
    \includegraphics[width=0.9\linewidth]{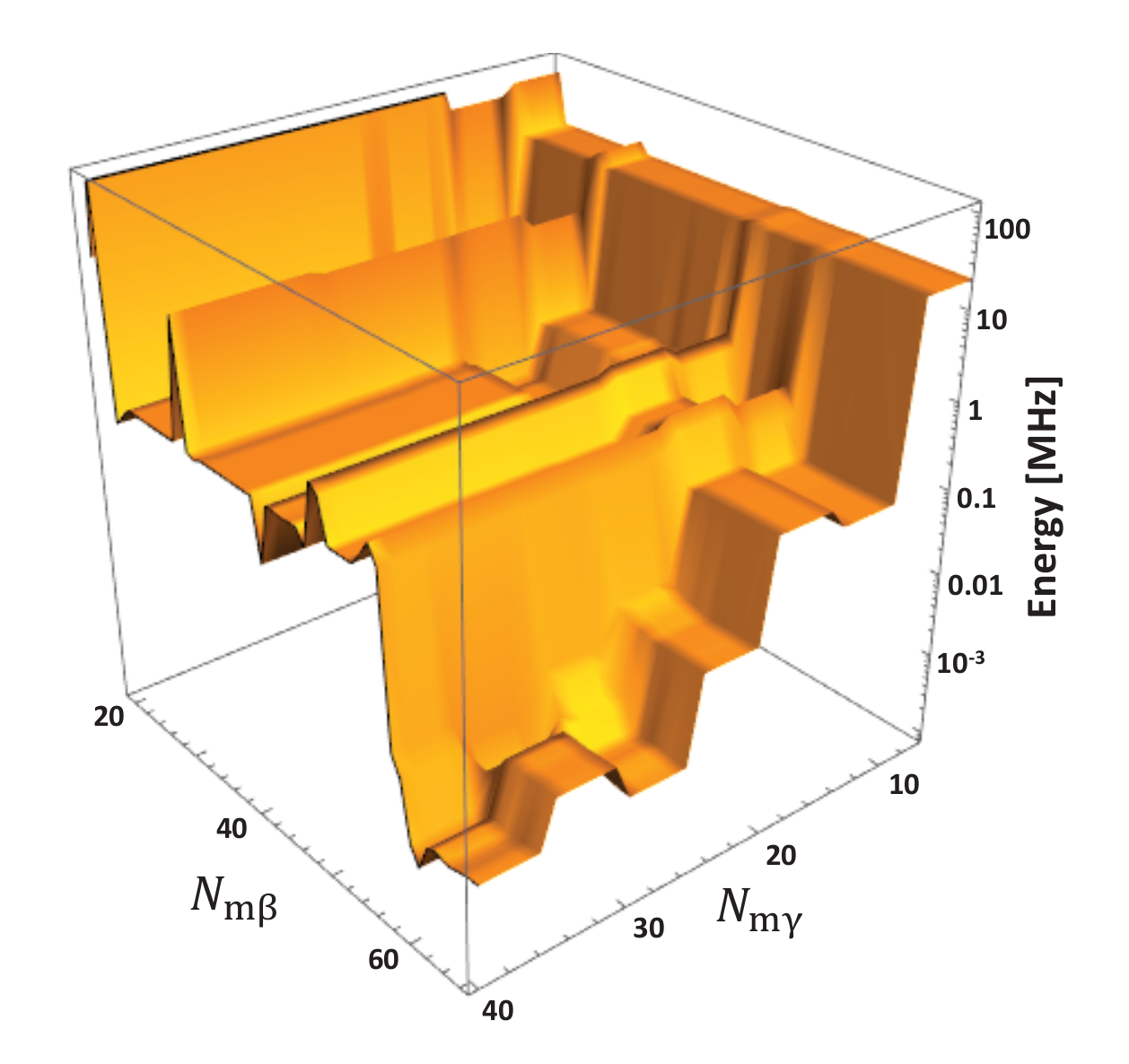}
    \caption[]
        {Errors in the calculated qubit energy splitting introduced by the subsystem truncation of figure ~\ref{fig:JPSQ_partition}. The horizontal axes are the number of eigenstates retained for subsystems $\beta$ and $\gamma$ prior to diagonalization of the total Hamiltonian in their product space. Sudden jumps in the error arise from low-lying excited states of high-energy  oscillator modes. }
        \label{fig:truncation_fig}
    \end{center}
\end{figure}


\noindent where three terms in $\hat{H}_{\beta\gamma}$ correspond to the three bold lines crossing between the two subsystems in fig.~\ref{fig:JPSQ_partition}. Recall that the total Hilbert space used to describe this circuit in fig.~\ref{fig:JPSQplot} above has dimension: $N_\textrm{tot}=(2q^\textrm{Jm}+1)\times(2q^\textrm{Im}+1)\times(\nu^\textrm{Om}_\ell+1)\times(\nu^\textrm{Om}_\textrm{p}+1)\times(\nu^\textrm{Om}_{\delta}+1)\times(\nu^\textrm{Om}_\textrm{R}+1)\times(\nu^\textrm{Om}_\textrm{L}+1)=39,600$. Using the same physical-layer truncation, the two subsystems defined in fig.~\ref{fig:JPSQ_partition}, however, have dimensions: $N_\beta=(2q^\textrm{Jm}+1)\times(\nu^\textrm{Om}_{\delta}+1)\times(\nu^\textrm{Om}_\textrm{R}+1)\times(\nu^\textrm{Om}_\textrm{L}+1)=495$ and $N_\gamma=(2q^\textrm{Im}+1)\times(\nu^\textrm{Om}_\textrm{p}+1)\times(\nu^\textrm{Om}_\ell+1)=80$. 

Figure ~\ref{fig:truncation_fig} shows the absolute error in the energy splitting between the two qubit levels of the total circuit (relative to the results of fig.~\ref{fig:JPSQplot}), as a function of $N_\beta$ and $N_\gamma$ used for the two subsystems when constructing and diagonalizing the total Hamiltonian of eq.~\ref{eq:Hs12}. A relative error at the $\sim$10 kHz level can be obtained using $N_\beta$=60, $N_\gamma$=25, for which the final Hamiltonian has a dimension of  $N_\beta\times N_\gamma$=1500. Although this matrix is much less sparse than those written in the physical basis of charge and oscillator levels, and constructing it requires several steps (evaluate and diagonalize the two subsystem Hamiltonians, re-express the subsytem interaction in the product basis, construct the new low-energy Hamiltonian for the full system, and diagonalize) we still find that this procedure is already more than an order of magnitude faster than the brute-force approach of section~\ref{s:JPSQ}, bringing the total calculation time from around a minute to only a few seconds. Although this is a significant improvement for this particular circuit, it is in fact not enough to go much farther in system size or complexity. The addition of only several more quantum degrees of freedom will immediately require either one or more of the subsystems to increase in size by a factor of tens or hundreds, or even worse, the addition of a third subsystem, causing the final (non-sparse) diagonalization to grow in size by such a factor. Either of these situations will immediately push to the edge of simulatability.

\section{Perturbative corrections for subsystem truncation}\label{s:truncation}

We now focus on the second of our two points above, the expected separation of energy scales, and how we can use it to improve the situation further. Consider the fact that while we are talking about calculating with high precision a qubit energy splitting of around ~1 GHz magnitude, the oscillator mode frequencies for this circuit [c.f., eq.~\ref{eq:HO}] range from $\sim$100 GHz to above 1 THz. Furthermore, if we examine the range of subsystem energies [c.f., eq.~\ref{eq:subeig}] that correspond to the Hilbert space truncation of $N_\beta$=60, $N_\gamma$=25, we find that this gives an energy range of 485 GHz for subsystem $\beta$ and 2.44 THz for subsystem $\gamma$: orders of magnitude larger than the scale of interest to us. To exploit this, we now describe a perturbative method to include the effects of these high-energy degrees of freedom, without increasing the size of the low-energy Hilbert space.

We formally divide the Hilbert space of each subsystem into low- and high-energy subspaces which we label g and e, using the following projection operator notation:

\begin{align}
\hat{\Pi}^\textrm{g}_\beta&\equiv\sum_{g=1}^{N_{\textrm{m}\beta}}|g_\beta\rangle\langle g_\beta|&\hat{\Pi}^\textrm{e}_\beta\equiv&\sum_{e>N_{\textrm{m}\beta}}|e_\beta\rangle\langle e_\beta| \nonumber\\
\hat{\Pi}^\textrm{g}&\equiv\prod_\beta\hat{\Pi}^\textrm{g}_\beta&\hat{\Pi}^\textrm{e}\equiv&\prod_\beta\hat{\Pi}^\textrm{e}_\beta \label{eq:PQproj}
\end{align}

\noindent The $|g_\beta\rangle$ subspace of subsystem $\beta$ extends up to the highest-energy level $N_{\textrm{m}\beta}$ (as before), and the $|e_\beta\rangle$ subspace may contain any chosen set of states orthogonal to $|g_\beta\rangle$. Note that unlike the typical case, here their union need not span the entire Hilbert space. Using this notation, we formally rewrite the Schr\"{o}dinger equation projected onto the $|g\rangle$ subspace:

\begin{equation}
\hat{\Pi}^\textrm{g}\left[E-\hat{H}_0-\hat{H}_1-\delta\hat{H}_1^\textrm{g}(E)\right]\hat{\Pi}^\textrm{g}|\Psi\rangle=0\label{eq:Hproj}
\end{equation}

\noindent where $\delta\hat{H}_1^\textrm{g}(E)$ is the correction to the $\Pi^\textrm{g}$ projection of the interaction Hamiltonian due to the presence of the $|e\rangle$ subspaces, which we can formally write as \footnote{The derivation of eq.~\ref{eq:resolvent} is closely related to, for example, the projection operator methods used in scattering calculations \cite{feshbach}, and adiabatic elimination of excited states \cite{sorensen}. As written, it is still exact if $\hat{\Pi}^\textrm{g}+\hat{\Pi}^\textrm{e}=\hat{I}$.}:

\begin{eqnarray}
\delta\hat{H}_1^\textrm{g}(E)=\hat{\Pi}^\textrm{g}\hat{H}_1\hat{\Pi}^\textrm{e}\frac{1}{E-\hat{H}_0-\hat{H}_1}\hat{\Pi}^\textrm{e}\hat{H}_1\hat{\Pi}^\textrm{g}\hspace{2.5cm}\label{eq:resolvent}\\
=\hat{\Pi}^\textrm{g}\hat{H}_1\left[\frac{\hat{\Pi}^\textrm{e}}{E-\hat{H}_0^\textrm{e}}+\frac{\hat{\Pi}^\textrm{e}}{E-\hat{H}_0^\textrm{e}}\hat{H}_1\frac{\hat{\Pi}^\textrm{e}}{E-\hat{H}^\textrm{e}_0}+...\right]\hat{H}_1\hat{\Pi}^\textrm{g}\nonumber
\end{eqnarray}

\noindent To make the structure of eq.~\ref{eq:resolvent} more apparent, we focus here on the linear electromagnetic interaction between subsystems $\beta$ and $\gamma$, writing the first two terms of eq.~\ref{eq:Veq}:

\begin{eqnarray}
\hat{H}_{\beta\gamma}&\equiv&\frac{1}{2}\hat{\vec{Q}}_\beta\cdot\mathbf{C}_{\beta\gamma}^{-1}\cdot\hat{\vec{Q}}_\gamma+\frac{1}{2}\hat{\vec{\Phi}}_\beta\cdot\mathbf{L}_{\beta\gamma}^{-1}\cdot\hat{\vec{\Phi}}_\gamma\nonumber\\
&&+\frac{1}{2}\hat{\vec{Q}}_\gamma\cdot\mathbf{C}_{\gamma\beta}^{-1}\cdot\hat{\vec{Q}}_\beta+\frac{1}{2}\hat{\vec{\Phi}}_\gamma\cdot\mathbf{L}_{\gamma\beta}^{-1}\cdot\hat{\vec{\Phi}}_\beta
\end{eqnarray}

\noindent The effective interaction in the $|g\rangle$ subspace can then be written:

\begin{widetext}
\begin{equation}
\hat{H}_{\beta\gamma}^\textrm{g}+\delta\hat{H}_{\beta\gamma}^\textrm{g}(E)\approx\hat{\Pi}^\textrm{g}_\beta\hat{\Pi}^\textrm{g}_\gamma\hat{H}_{\beta\gamma}\hat{\Pi}^\textrm{g}_\beta\hat{\Pi}^\textrm{g}_\gamma+\hat{\Pi}^\textrm{g}_\beta\hat{\Pi}^\textrm{g}_\gamma\hat{H}_{\beta\gamma}\left[\frac{\hat{\Pi}^\textrm{g}_\beta\hat{\Pi}^\textrm{e}_\gamma}{E-\hat{H}^\textrm{g}_\beta-\hat{H}^\textrm{e}_\gamma}+\frac{\hat{\Pi}^\textrm{e}_\beta\hat{\Pi}^\textrm{g}_\gamma}{E-\hat{H}^\textrm{e}_\beta-\hat{H}^\textrm{g}_\gamma}+\frac{\hat{\Pi}^\textrm{e}_\beta\hat{\Pi}^\textrm{e}_\gamma}{E-\hat{H}^\textrm{e}_\beta-\hat{H}^\textrm{e}_\gamma}\right]\hat{H}_{\beta\gamma}\hat{\Pi}^\textrm{g}_\beta\hat{\Pi}^\textrm{g}_\gamma
\label{eq:dyson}
\end{equation}
\end{widetext}

\noindent Note that the $E$ appearing in eqs.~\ref{eq:Hproj}-\ref{eq:dyson} is the total energy, and to apply this perturbation theory it must be approximated in the $|g\rangle$ subspace of the two subsystems (typically by assuming that energy differences within $|g\rangle$ are much smaller than the energy defect to important states in $|e\rangle$). In general, this second-order perturbation theory can of course begin to break down if states in $|g\rangle$ approach the energy range of the $|e\rangle$ subspace. Conversely, it is well-suited to the situation typically encountered in circuits of interest here, where there is a large separation of energy scales (and thus a natural energy at which to place the boundary between the $|g\rangle$ and $|e\rangle$ subspaces).

The first term in eq.~\ref{eq:polarize} is the direct (unperturbed) interaction in the $|g\rangle$ subspace, the second (third) terms involve virtual excitation of $\beta$ ($\gamma$) into its excited $|e\rangle$ subspace, and the last term involves virtual excitation of both subsystems. To emphasize the physical intuition underlying this structure, we can rewrite eq.~\ref{eq:dyson} thus:

\begin{widetext}
\begin{equation}
\hat{H}_{\beta\gamma}^\textrm{g}+\delta\hat{H}_{\beta\gamma}^\textrm{g}(E)\approx\vec{\mathcal{F}}^\textrm{gg}_\beta\cdot\mathbf{\mathcal{M}}_{\beta\gamma}\cdot\vec{\mathcal{F}}^\textrm{gg}_\gamma
+\frac{1}{2}\vec{\mathcal{F}}^\textrm{gg}_\beta\cdot\left(\mathbf{\mathcal{M}}_{\beta\gamma}\cdot\hat{\mathcal{A}}_\gamma\cdot\mathbf{\mathcal{M}}_{\gamma\beta}\right)\cdot\vec{\mathcal{F}}^\textrm{gg}_\beta
+\frac{1}{2}\vec{\mathcal{F}}^\textrm{gg}_\gamma\cdot\left(\mathbf{\mathcal{M}}_{\gamma\beta}\cdot\hat{\mathcal{A}}_\beta\cdot\mathbf{\mathcal{M}}_{\beta\gamma}\right)\cdot\vec{\mathcal{F}}^\textrm{gg}_\gamma+\hat{U}^\textrm{d}_{\beta\gamma}\hspace{0.7cm}
\label{eq:polarize}
\end{equation}
\end{widetext}

\noindent such that the first terms appears as a static interaction between $|g\rangle$-subspace dipoles of the two subsytems, the second and third terms describe polarization of one system by the static dipole moments of the other, and the final term is a dispersion interaction between fluctuation dipoles of the two subsystems. We have used the following definitions:

\begin{equation}
\vec{\mathcal{F}}^\textrm{ab}_\beta\equiv\begin{pmatrix}
      \hat{\vec{Q}}^\textrm{ab}_\beta \\
      \hat{\vec{\Phi}}^\textrm{ab}_\beta
\end{pmatrix},
\textrm{a,b}\in\{\textrm{g,e}\},\hspace{0.15cm}
\hat{\vec{O}}^\textrm{ab}_\beta\equiv\hat{\Pi}^\textrm{a}_\beta\hat{\vec{O}}_\beta\hat{\Pi}^\textrm{b}_\beta,\;O\in\{Q,\Phi \}
\end{equation}

\begin{equation}
\mathbf{\mathcal{M}}_{\beta\gamma}\equiv\begin{pmatrix}
      \mathbf{C}_{\beta\gamma}^{-1} & 0\\
      &\\
      0 & \mathbf{L}_{\beta\gamma}^{-1}
\end{pmatrix}\hspace{0.5cm}
\hat{\mathcal{A}}_\beta\equiv\begin{pmatrix}
      \hat{\alpha}^{QQ}_\beta & \hat{\alpha}^{Q\Phi}_\beta\\
      &\\
      \hat{\alpha}^{\Phi Q}_\beta & \hat{\alpha}^{\Phi\Phi}_\beta
\end{pmatrix}
\end{equation}

\begin{eqnarray}
\hat{\mathbf{\alpha}}^{OP}_\beta&\equiv&\hat{\vec{O}}^\textrm{ge}_\beta\frac{1}{E-\hat{H}^\textrm{g}_\gamma-\hat{H}^\textrm{e}_\beta}\hat{\vec{P}}^\textrm{eg}_\beta,\hspace{0.5cm}O,P\in\{Q,\Phi \}\nonumber\\
\hat{U}^\textrm{d}_{\beta\gamma}&\equiv&\frac{\left(\vec{\mathcal{F}}^\textrm{ge}_\beta\cdot\mathbf{\mathcal{M}}_{\beta\gamma}\cdot\vec{\mathcal{F}}^\textrm{eg}_\gamma\right)\left(\vec{\mathcal{F}}^\textrm{ge}_\gamma\cdot\mathbf{\mathcal{M}}_{\gamma\beta}\cdot\vec{\mathcal{F}}^\textrm{eg}_\beta\right)}{E-\hat{H}^\textrm{e}_\gamma-\hat{H}^\textrm{e}_\beta}\hspace{0.25cm}
\end{eqnarray}

\noindent where the quantity $\vec{\mathcal{F}}^\textrm{gg}_\beta$ plays the role of a static dipole moment in the $|g\rangle$ subspace of subsystem $\beta$, $\mathbf{\mathcal{M}}_{\beta\gamma}$ describes the coupling between subsystems $\beta$ and $\gamma$, $\hat{\mathcal{A}}_\beta$ is a generalized polarizability of subsystem $\beta$, and $\hat{U}^\textrm{d}_{\beta\gamma}$ is the dispersion interaction between subsystems $\beta$ and $\gamma$. Note that the 2-entry matrix-vector notation here does not refer to the two subsystems, but to electric and magnetic interactions. Although one does not normally think of electric and magnetic interactions coupling to each other, in the present case this can occur when the oscillator modes exhibit Josephson nonlinearity. For example, the operator $\hat{\alpha}^{Q\Phi}_\beta$ describes an interaction between electric and magnetic dipole moments induced in subsystem $\beta$ by the static moments of subsystem $\gamma$. Similarly, $\hat{U}^\textrm{d}_{\beta\gamma}$ in general involves interactions between electric and magnetic quantum fluctuations of the two subsystems.


Before proceeding, we highlight that the approximation being made here is associated purely with truncation of the subsystem Hilbert spaces. That is, we are \textit{not} treating the subsystem interactions themselves perturbatively, but rather only the \textit{low-energy corrections to them that arise from high energy states}. Once we construct an effective interaction between subsystems in their $|g\rangle$ subspaces, the resulting Hamiltonian is then diagonalized exactly, thereby including the effective interaction to all orders.

\begin{figure}[h!]
    \begin{center}
    \includegraphics[width=1.0\linewidth]{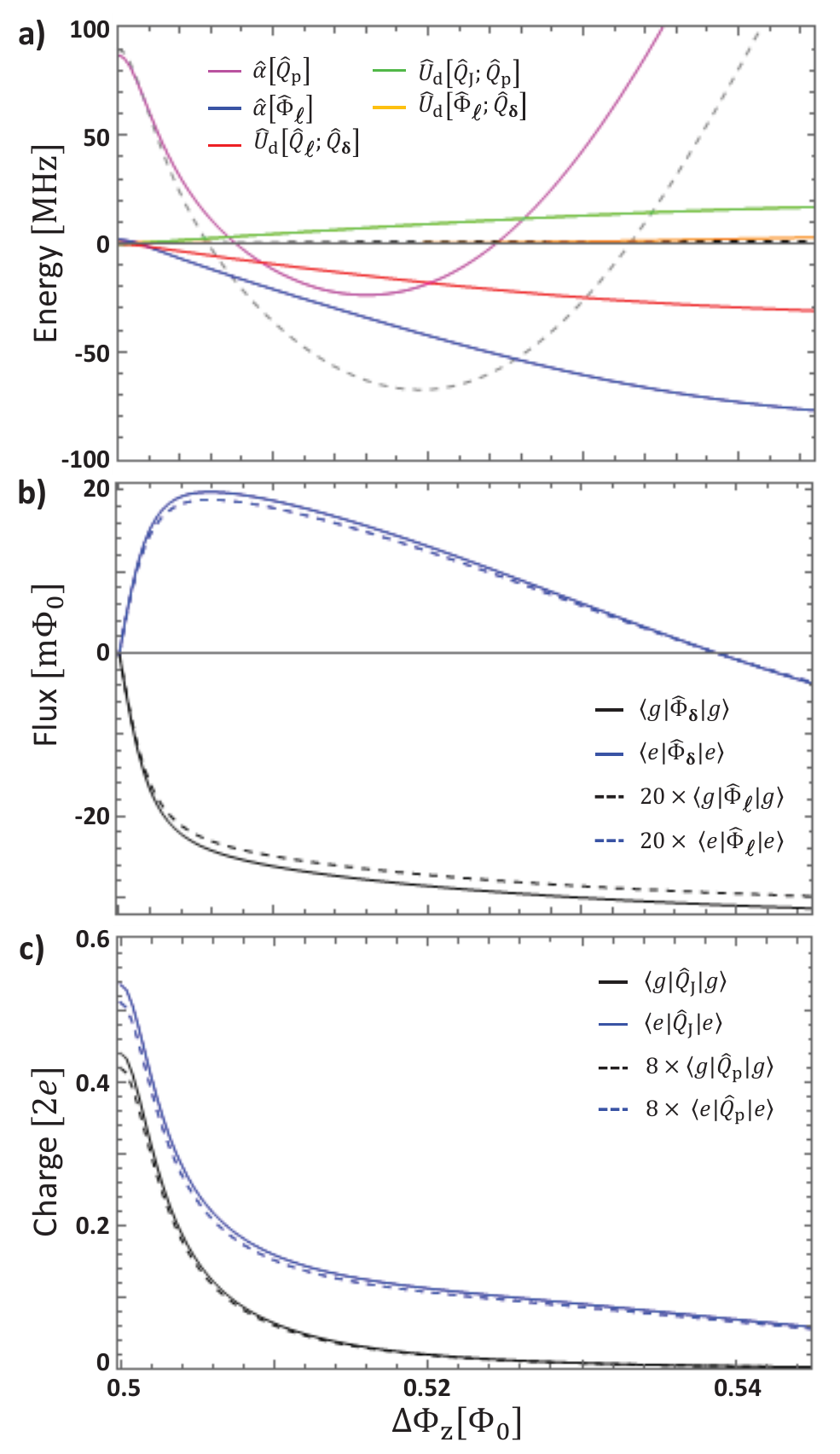}
    \caption[]
        {Truncation corrections for the Josephson phase-slip qubit circuit of fig.~\ref{fig:JPSQ}. Panel (a): corrections for the flux sweep of fig.~\ref{fig:JPSQplot}(b), with $Q_g=0.45$ (for which the qubit splitting is 1.6 GHz at the flux symmetry point $\Phi_z=\Phi_0/2$). The gray and black dashed lines show the error in the qubit splitting with and without truncation corrections. The colored solid lines show the non-negligible contributions to this correction, arising from the interaction of low energy states with much higher energies via the inter-subsystem interaction of eq.~\ref{eq:JPSQHsub}. Panels (b) and (c) show the expectation values of the nonzero dipole moments for the ground (black) and excited (blue) states. Solid (dashed) lines are static (induced) dipoles.}
        \label{fig:polarization_fig}
    \end{center}
\end{figure}

It may occur to the astute reader, noting that these perturbative corrections are most appropriate for interactions between high-frequency oscillator modes, that it is in fact always possible to transform the canonical oscillator mode coordinates to a set of modes that are decoupled from each other \cite{cottet} (these are simply the noninteracting, normal modes of the linear part of the circuit). Such a transformation by definition removes all (or a desired subset of) the couplings between oscillator modes, thereby obviating the need for polarization corrections of the kind just described. Unfortunately, this produces an unintended numerical consequence in circuits with Josephson nonlinearity: transforming to normal modes of the oscillators invariably results in many more interactions between them and the Josephson modes, which appear in the form of oscillator charge displacement operators as in eq.~\ref{eq:Ldisp}. The non-sparsity of these matrices (in comparison to the original charge and flux operators of eqs.~\ref{eq:phiL}-\ref{eq:QL}, which are band diagonal), strongly decreases the sparsity of the overall numerical Hamiltonian, and in fact tends to increase the diagonalization time well beyond that required to implement the corrections perturbatively. Furthermore, such a transformation produces the added complication of making the modes themselves dependent on the numerical input parameters, and thereby requiring often complex basis transformations in order to make direct comparisons between the eigenstates obtained for different parameters.


Figure \ref{fig:polarization_fig}(a) shows the results obtained by using this perturbative procedure to reproduce the calculation of fig.~\ref{fig:JPSQplot}(b). The vertical axis shows, similar to fig.~\ref{fig:truncation_fig}, the error in the qubit energy splitting (between the first two energy eigenstates) relative to the result obtained by diagonalizing the full matrix of dimension $39,600$ and shown in fig.~\ref{fig:JPSQplot}(b). However, this time, after diagonalizing the $\beta$ and $\gamma$ subsystem Hamiltonians (having dimensions of 495 and 80 as above in section ~\ref{s:partition}) the final Hamiltonian was diagonalized using a trunctation of only $N_\beta$=6, $N_\gamma$=5 (corresponding to a physically-relevant energy range of $\sim h\times$50 GHz above the ground state of each subsystem), having a dimension: $N_\beta\times N_\gamma$=30. The dashed gray line in the figure shows the error in this case, which is not only quite large, but depends non-monotonically on the flux bias. This nonlinear behavior with flux is a result of a broad avoided crossing with the excited qubit level, evident in fig.~\ref{fig:JPSQplot}(b) at around $\Phi_z\sim0.528\Phi_0$, which causes it to deviate from the nearly linear behavior of a persistent current state, and bend downward. The dashed black line in fig.~\ref{fig:polarization_fig}(a) shows the result when we apply the perturbative corrections described above in this section, where the g and e spaces are taken, respectively, as levels 1-6 and 7-60 for subsystem $\beta$, and 1-5 and 6-25 for subsystem $\gamma$, such that in total they span the same subspaces mentioned in section~\ref{s:partition} above. The perturbative result is barely distinguishable from the horizontal axis, with an error $<$1 MHz over the flux range plotted, and only $\sim$20 kHz at $\Delta\Phi_z=0.505\Phi_0$, the flux used in fig.~\ref{fig:truncation_fig}. The additional solid lines illustrate the contributions of the various components of the truncation error, by showing the error when individual correction terms are removed from the calculation. From the interaction Hamiltonian from eq.~\ref{eq:JPSQHsub}, we have contributions from six polarizabilities: $\hat\alpha[\hat{Q}_\textrm{p}]$, $\hat\alpha[\hat{Q}_\textrm{J}]$, $\hat\alpha[\hat{Q}_{\ell}]$, $\hat\alpha[\hat{Q}_\delta]$, $\hat\alpha[\hat\Phi_{\ell}]$, and $\hat\alpha[\hat\Phi_\delta]$ and three dispersion interactions: $\hat{U}_\textrm{d}[\hat\Phi_{\ell};\hat\Phi_\delta]$, $\hat{U}_\textrm{d}[\hat{Q}_\textrm{J};\hat{Q}_\textrm{p}]$, and $\hat{U}_\textrm{d}[\hat{Q}_{\ell};\hat{Q}_\delta]$.

The two largest of these, associated with $\hat\alpha[\hat{Q}_\textrm{p}]$ and $\hat\alpha[\hat\Phi_{\ell}]$, are shown in \ref{fig:polarization_fig}(a) with magenta and blue lines, respectively. The reason these two are the largest contributions can be understood from panels (b) and (c), where solid lines show the expectation values, for the ground state (black) and first excited state (blue), of the four nonzero static dipole moments \footnote{Note that these quantities are evaluated using the full simulation of fig.~\ref{fig:JPSQplot}.}. The solid lines in panel (b) show the expectation value of the loop flux operator $\hat\Phi_\delta$, which corresponds to the qubit persistent current. As expected, the expectation values for the ground and excited states start at zero at the symmetry point $\Phi_z=\Phi_0/2$, and increase equally in opposite directions. For an ideal persistent-current qubit, these two quantities would saturate at approximately equal and opposite values far from $\Phi_z=\Phi_0/2$; however, the excited state deviates from this behavior here due to the strong avoided crossing mentioned above. The dashed lines show the expectation value (magnified by 20 times) of $\hat\Phi_{\ell}$, the additional loop oscillator mode to which $\hat\Phi_\delta$ is coupled by the inter-subsystem interaction of eq.~\ref{eq:JPSQHsub}. The key feature of this plot is that the two expectation values have identical shapes, leading to the conclusion that the strong (A subsystem) static dipole $\hat\Phi_\delta$ associated with the persistent current \textit{magnetizes} the B subsystem oscillator mode flux $\hat\Phi_{\ell}$. The additional magnetization energy associated with this inter-subsystem coupling is responsible for the correction indicated with a blue line in panel (a). Similarly, panel (c) shows with solid lines the large static electric dipole associated with $\hat{Q}_\textrm{J}$, the mode polarized by the island offset charge $\Delta Q_I$, while the dashed lines then show the induced polarization of $\hat{Q}_\textrm{p}$ (magnified 8 times), the B subsystem mode charge to which it is coupled by the inter-subsystem interaction of eq.~\ref{eq:JPSQHsub}. The electric polarization energy associated with this coupling is responsible for the correction indicated with a magenta line in panel (a). The strongly non-monotonic shape of this curve, as mentioned above, is due to a strong avoided crossing with the excited qubit level, which modifies its effective polarizability. The three additional solid lines in (a) arise from the three possible dispersion interactions.

For even larger systems, the partitioning and diagonalization process described by eqs.~\ref{eq:subsys}-\ref{eq:subops} can be repeated iteratively, at each step diagonalizing a new set of (fewer, and larger) subsystems, and then re-expressing the interactions between these subsystems in the resulting eigenbasis, before diagonalizing again. An example of the results of this procedure is contained in ref. ~\cite{JPSQ}, where the circuit of fig. 10 used four successive steps of diagonalization. The perturbative truncation corrections of section ~\ref{s:truncation} can also be used effectively at later stages of diagonalization, though it will typically require detailed investigation of the subsystems to be sure that the appropriate excited states are included to calculate the corresponding polarizabilities. In many cases, with judicious choice of the subsystem partitioning, all important polarizabilites can be captured at the first level of partitioning described in section ~\ref{s:truncation}, after which they need only be re-expressed in the new eigenbases generated at each subsequent level until the interactions they capture are contained within a single subsystem and are then included in its diagonalization.
 

\section{Conclusion}\label{s:conclusion}

We have described a new theoretical and numerical framework for simulating the static properties of quantum superconducting circuits, whose purpose is to substantially broaden the reach of possible new circuit designs beyond the typical, simple qubit circuits currently in widespread use. This capability not only allows for detailed, predictive simulation of larger and more complex combinations of existing qubits, but also for the design of entirely new kinds of quantum circuits which could not previously be considered due to the difficulty of the required design simulations.

The logical next step beyond this work is its extension to dynamics simulations in the instantaneous adiabatic eigenbasis \cite{kerman,MRFS,Albash2012,vinciNS}, work which is currently underway.

\section{Acknowledgments}

This research was funded by the Ofﬁce of the Director of National Intelligence (ODNI), Intelligence Advanced Research Projects Activity (IARPA), and by the Assistant Secretary of Defense for Research, Engineering under Air Force Contract No. FA8721-05-C-0002. The views and conclusions contained herein are those of the authors and should not be interpreted as necessarily representing the ofﬁcial policies or endorsements, either expressed or implied, of ODNI, IARPA, or the US Government.




\bibliography{JJLsim21}


\end{document}